\documentclass[
 amssymb, amsmath,
 reprint,%
aps, prl,
superscriptaddress,%
floatfix
]{revtex4-1}

\usepackage{graphicx}
\usepackage[colorlinks=true,linkcolor=blue]{hyperref}

\usepackage[T1]{fontenc} 
\usepackage[utf8]{inputenc}
\usepackage{lmodern}

\usepackage{subcaption}
\usepackage{etoolbox}
\usepackage[separate-uncertainty=true]{siunitx}
\usepackage{upgreek}
\usepackage{chemmacros}

\chemsetup{
 formula = mhchem,
greek = chemgreek,
 modules = thermodynamics,
 modules = redox,
 modules = reactions
}

\renewrobustcmd{\bfseries}{\fontseries{b}\selectfont}
\renewrobustcmd{\boldmath}{}

\newcommand*{\Hform}{$\enthalpy*(f){}{}$}
\newcommand*{\Tref}{\SI{298.15}{\kelvin}}
\newcommand*{\Tzero}{\SI{0}{\kelvin}}
\newcommand{\etal}{et al.}
\newcommand{\dcm}{$\mathrm{^2DC^M}$}

\begin{document}
\author{Sophie Kervazo}
\affiliation{Univ. Lille, CNRS, UMR 8523 - PhLAM - Physique des Lasers Atomes et Molécules, F-59000 Lille, France}
\altaffiliation{Department of Chemistry and Chemical Biology, McMaster University,  1280 Main Street West, Hamilton L8S 4M1, Canada}
\author{Florent Réal}
\affiliation{Univ. Lille, CNRS, UMR 8523 - PhLAM - Physique des Lasers Atomes et Molécules, F-59000 Lille, France}
\author{François Virot}
\affiliation{Institut de Radioprotection et de Sûreté Nucléaire (IRSN) PSN-RES, Cadarache, Saint Paul Lez Durance 13115, France}
\author{André Severo Pereira Gomes}
\affiliation{Univ. Lille, CNRS, UMR 8523 - PhLAM - Physique des Lasers Atomes et Molécules, F-59000 Lille, France}
\author{Valérie Vallet}
\affiliation{Univ. Lille, CNRS, UMR 8523 - PhLAM - Physique des Lasers Atomes et Molécules, F-59000 Lille, France}
\email{valerie.vallet@univ-lille.fr}

\title[Accurate Predictions of Volatile Plutonium Thermodynamic Properties]
  {Accurate Predictions of Volatile Plutonium Thermodynamic Properties}

\begin{abstract}
The ability to predict the nature and amounts of plutonium emissions in industrial
accidents, such as in solvent fires at PUREX nuclear reprocessing facilities, is a key concern of nuclear safety agencies. In accident conditions and in the presence of oxygen and water vapor, plutonium is expected to form the three major volatile species \ce{PuO2}, \ce{PuO3}, and \ce{PuO2(OH)2}, for which the thermodynamic data  necessary for predictions (enthalpies of formation and heat capacities) presently show either large uncertainties or are lacking. In this work we aim to alleviate such shortcomings by obtaining the aforementioned data via relativistic correlated electronic structure calculations employing the multi-state complete active apace with second-order perturbation theory (MS-CASPT2) with state-interaction RASSI spin-orbit coupling approach, which is able to describe the multireference character of the ground-state wave functions of \ce{PuO3} and \ce{PuO2(OH)2}. We benchmark this approach by comparing it to relativistic coupled cluster calculations for the ground, ionized, and excited states of \ce{PuO2}. Our results allow us to predict enthalpies of formation {\Hform}({\Tref}) of \ce{PuO2}, \ce{PuO3}, and \ce{PuO2(OH)2} to be \num{-449.5\pm8.8}, \num{-553.2\pm27.5}, and \SI{-1012.6\pm38.1}{\kJ\per\mol}, respectively, which confirm the predominance of plutonium dioxide, but also reveal the existence of plutonium trioxide in the gaseous phase under oxidative conditions, though the partial pressures of \ce{PuO3} and \ce{PuO2(OH)2} are nonetheless always rather low under a wet atmosphere. Our calculations also permit us to reassess prior results for \ce{PuO2}, establishing that the ground state of the \ce{PuO2} molecule is mainly of $\mathrm{^{5}\Sigma_{g}^+}$ character, as well as to confirm the experimental value for the adiabatic ionization energy of \ce{PuO2}.
\end{abstract}
\maketitle

\section{Introduction}\label{sec:introduction}

A key step in the reprocessing of spent nuclear fuel is the PUREX (plutonium uranium refining by extraction) liquid-liquid process, through which uranium and plutonium are separated from the minor actinides. Thanks to its remarkable selectivity and the stability properties, tri-n-butyl phosphate (TBP) is used as the organic extractant. However, its high viscosity and density impose that the organic phase should be diluted within a hydrocarbon solvent. In the French La Hague reprocessing plant, hydrogenated tetrapropylene (HTP) is currently used as a TBP diluent. As HTP is a highly flammable substance, this could, in case of an accidental event, trigger a fire of the uranium-plutonium containing organic solvent and may thus induce a release in the environment and atmosphere of highly radioactive species. The knowledge of their chemical behaviors is thus mandatory to carry out safety and risk analysis. 

The presence of plutonium induces considerable difficulties to perform experimental investigations of solvent fires. One way to overcome these is to use a plutonium surrogate in order to carry out fire experimental campaigns. To make a selection of surrogate candidates, their behavior should be similar in all oxidation states. Considering for example the gaseous state, the volatility of Pu(VI), which still remains an open issue, should be addressed specifically because some of the candidates (Th, Ce) \cite{actinide-Sulka-JPC2014-118-10073} cannot extend their oxidation degree up to VI. An alternative to surrogates is the development of a predictive model which includes all relevant physical and chemical transformations of plutonium-containing substances. Whatever the case, both approaches need reliable and accurate molecular properties of the gaseous species involved in the Pu+O+H system, required to calculate their thermodynamic properties. 

The thermodynamic properties of the Pu+O+H system have been widely studied and assessed (see refs \citenum{actinide-Konings-JPCRD2014-43-013101} and \cite{actinide-Gueneau-JNM2008-378-257} and the references therein). Nonetheless, some uncertainties remain in the thermodynamic properties of some species: for the gaseous plutonium dioxide, \ce{PuO2}, the assessment of its standard heat of formation reveals a discordance between second and third law analyses of experimental data \cite{actinide-Konings-JPCRD2014-43-013101}. In addition, experimental data (and in particular thermodynamic data) for other Pu(VI) molecules, such as \ce{PuO2(OH)2} and \ce{PuO3}, are scarce, and their existence in the gaseous phase remains an open question. Ronchi~{\etal} \cite{Ronchi-JNM2000} have carried out mass spectrometric measurements of effusing vapor over plutonium oxide sample where \ce{PuO3(g)}  has been detected but only as a nonequilibrium species. Then, transpiration experiments of plutonium oxide have been performed by Krikorian~{\etal} \cite{actinide-Krikorian-JNM1997-247-161} under oxygen as well as a mixture of oxygen and steam. From analogies with gaseous uranium species, the authors considered \ce{PuO2(OH)2} and \ce{PuO3} molecules as the effusing species. Hübener~{\etal} \cite{actinide-Hubener-RA2008-96-781} have studied by thermochromatography the volatility of plutonium oxide under wet oxygen flux, and their results highlighted the formation of volatile Pu(VI) expected as \ce{PuO2(OH)2}. 

To predict the respective quantities of the three target volatile products of plutonium in the presence of oxygen or steam, accurate thermal functions (heat capacity, entropy, enthalpy increment) and enthalpies of formation are needed. For gas-phase molecules, this all boils down to having data on their electronic structure in order to directly calculate the vibrational-rotational partition functions. The electronic partition functions in turn require the knowledge of all the low-lying excited states within \SI{10000}{\per\cm} and their spectroscopic labeling as degeneracies enter the statistical weights. If not known from direct spectroscopic measurements, these are often roughly estimated from chemically and spectroscopically analogous systems, such as crystals or molecules containing another actinide. However, the investigation of plutonium-containing molecules with quantum chemical methods remains a big challenge because they exhibit complex electronic structures, and as such it is often of little use to estimate their properties from chemical analogues. The presence of open 5f, 6d, and 7s orbitals on the plutonium atom leads to dense valence electronic spectra, in which the numerous degenerate or quasi-degenerate electronic states are suspected to have strong multireference characters. 

A potential route for  handling electron correlation and relativistic effects simultaneously for multireference heavy element complexes is to work in the four-component (4C) framework, with the Dirac-Coulomb, Dirac-Coulomb-Gaunt, or Dirac-Coulomb-Breit Hamiltonian or with the exact two-component (X2C) approaches using a molecular mean-field (MMF) approach \cite{Sikkema2009} ({\dcm}). The multireference intermediate Hamiltonian Fock-space coupled cluster (IHFSCC) has been successfully applied to actinide-containing molecules \cite{actinide-Infante-JCP2006-125-074301,actinide-Infante-JCP2007-127-124308}, though with available implementations it is not possible to treat states differing by more than two electrons from a closed-shell reference. This is problematic  for \ce{PuO3}, which turns out to have a complex multireference ground-state with significant contributions from configurations with two or even four unpaired electrons \cite{actinide-Boguslawski-PCCP2017-19-4317,actinide-Kovacs-JMS2017-1132-95,actinide-Gomes-PCCP2008-10-5353,actinide-Ruiperez-JPC2009-113-1420,actinide-Weigand-JPC2009-113-11509,actinide-Real-JPC2009-113-12504,actinide-Tecmer-PCCP2011-13-6249,actinide-Tecmer-JCP2012-137-084308,actinide-Gomes-PCCP2013-15-15153}. On the other hand, at the fully relativistic level, \ce{PuO2} turns out to have a closed-shell ground state, allowing us to also estimate its standard enthalpy of formation with coupled-cluster single and double with perturbative triples ({\dcm}-CCSD(T)) calculations. The recent implementation of the fully relativistic EOM-CCSD method \cite{CCmethods-Shee-JCP2018-147-174113} offers the possibility to accurately compute the ionization potential (IP) of the \ce{PuO2} molecule, and new arguments are proposed to clarify why the experimental values reported by Rauh~{\etal} \cite{IP-Rauh-JCP1974-60-1396-1400,IP-Rauh-JCP1974-64-1862} and Capone~{\etal} \cite{actinide-Capone-JPC1999-103-10899} differ by about \SI{3}{\electronvolt} from the IP value measured by Santos~{\etal} \cite{IP-Santos-JPC2002-106-7190}. 

Considering {\dcm}-CCSD(T) to be the gold standard -- it was shown by Shee~{\etal}\cite{CCmethods-Shee-JCP2018-147-174113} that {\dcm} Hamiltonian yields results which are nearly indistinguishable from the Dirac–Coulomb Hamiltonian -- and the fact that there are more experimental data estimating the enthalpy of formation, the \ce{PuO2} molecular system represents an excellent candidate to benchmark relativistic multireference methods with an \emph{a posteriori} treatment of the spin-orbit coupling (SOC) interaction, namely the state-interaction-multistate second-order perturbation theory RASSI-MS-CASPT2 approach, hereafter referred to as SO-CASPT2, which has been used to compute the thermodynamic data of \ce{PuO3} and \ce{PuO2(OH)2}. High accuracy is achieved by extrapolating the electronic energies to the complete basis set limit. 

The details of the calculations are described in the following section, followed by a discussion of the electronic structure of the three aforementioned molecules. We continue with an analysis of the thermodynamic properties and a comparison to available experimental data and propose new heat capacity functions and standard entropies and enthalpies of formation. Finally, the volatility of Pu(VI) species is investigated within the framework of thermodynamic equilibrium calculations. The article closes with conclusions and perspectives for nuclear safety related issues.

\section{Computational Details}
\label{Sec:comp}
\subsection{Chemical Reactions Used to Derive the Standard Enthalpies of Formation of \ce{PuX} Molecules}
To determine the enthalpies of formation, {\Hform}({\Tref}), of the different plutonium oxides and oxyhydroxides, we consider the reactions presented in Table~\ref{Tab:reaction}. We begin by computing the enthalpies of the reactions presented in Table~\ref{Tab:reaction}, as:
\begin{align}\label{eq:1}
\enthalpy*(r){}{} =  \displaystyle\sum_{i} \left(E_i  + ZPE_{i} + \left[\state[pre=]{H}(\Tref)-\state[pre=]{H}({\Tzero})\right]_{i}\right),
\end{align}
that is by summing up electronic energies $E_i$, the vibrational zero point energies ($ZPE_i$) and the enthalpy increment, $\left[\state[pre=]{H}(\Tref)-\state[pre=]{H}({\Tzero})\right]_{i}$ at {\Tref} (see Table~S1) for the products, minus that of the reactants. For a generic reaction leading to the formation of \ce{PuX} (\ce{Pu + A -> PuX + B}), the standard enthalpy of formation, \enthalpy*(f){}({\Tref}), is computed as
\begin{align}\label{eq:2}
 \enthalpy*(f){}{}(PuX) = \enthalpy*(r){}{}  - \enthalpy*(f){}{} (B) \\+ \enthalpy*(f){}{}(Pu) +  \enthalpy*(f){}{}(A),
\end{align}
using the known enthalpies of formations of the \ce{Pu} and \ce{A} reactants and the \ce{B} product listed with their uncertainties in Table~S1 in the Supporting Information \cite{thermo-Ruscic-2019-ATcT-1.122e,actinide-Lemire-ChemThermo-2001}. Note that, because of the lack of thermodynamics data on plutonium species, it is not possible to use reactions that are isogyric or isodesmic, which might alter the final accuracy of our results; the measure of the resulting uncertainties within the 95\% confidence limit obtained for the average of these reactions is computed as explained in the Supporting Information, and reported in Table~\ref{Tab:PuOx_Hform}.

\begin{table*}[htp]
\begin{scriptsize}
\caption{Reactions Used to Derive the Standard Enthalpies of Formation of \ce{PuO2}, \ce{PuO3}, and \ce{PuO2(OH)2}}
\label{Tab:reaction}
\begin{tabular}{lccc}
\hline
& \ce{PuO2} & \ce{PuO3} & \ce{PuO2(OH)2} \\
\hline 
R$_{1}$ & \ce{Pu + O2 ->  PuO2} & \ce{Pu + 3/2 O2 -> PuO3} &  \ce{Pu + O2 + 2 H2 -> PuO2(OH)2} \\
R$_{2}$ & \ce{Pu + 2 O -> PuO2} & \ce{Pu + 3 O -> PuO3} & \ce{Pu + 4 O + 2 H -> PuO2(OH)2} \\
R$_{3}$ & \ce{Pu + H2O2 -> PuO2 +  H2} &  \ce{Pu + 3/2 H2O2 -> PuO3 + 3/2 H2O} & \ce{Pu + 3 H2O2 -> PuO2(OH)2 + 2 H2O}\\
R$_{4}$ & \ce{Pu + 2 OH ->  PuO2 + 2 H} &  \ce{Pu + 3 OH -> PuO3 + 3 H} &  \ce{Pu + 4 OH -> PuO2(OH)2 + 2 H} \\
R$_{5}$ & \ce{Pu + 2 H2O -> PuO2 +  2 H2} & \ce{Pu + 3 H2O -> PuO3 + 3 H2} &  \ce{Pu + 4 H2O -> PuO2(OH)2 + 2 H2} \\
R$_{6}$ & \ce{Pu + 2 H2O -> PuO2 +  4 H} & \ce{Pu + 3 H2O -> PuO3 + 6 H} &  \ce{Pu + 4 H2O -> PuO2(OH)2 + 6 H} \\
\hline
\end{tabular}
\end{scriptsize}
\end{table*}

\subsection{Two-Step Relativistic Calculations} 
\subsubsection{Geometries and Enthalpy Corrections}\label{comp:geometries}

The geometries were obtained from scalar relativistic DFT calculations, except for \ce{PuO2}, for which the optimal {\dcm}-CCSD(T) geometry was used. In the DFT calculations, the plutonium atom was described by a relativistic effective core potential (RECP) ECP60MWB \cite{ecp-Kuchle-JCP1994-100-7535} with the corresponding basis set of quadruple-$\zeta$ quality \cite{basis-Cao-JCP2003-118-487}, while for the lighter atoms augmented triple $\zeta$ (aug-cc-pVTZ) \cite{basis-Dunning-JCP1989-90-1007} basis sets were used. All of the DFT calculations were carried out with the B3LYP functional \cite{dft-Becke-JCP1993-98-5648} and the GAUSSIAN09 \cite{prog-G09} package. The ZPE values and enthalpy corrections needed in eq.~\ref{eq:1} were calculated from the harmonic vibrational frequencies. Anharmonic corrections to the vibrational partition functions were not included, as these would contribute only a few tenths of a \si{\kJ\per\mol} at most and only or those species involving hydrogen atoms. Using the vibrational perturbation theory\cite{spectro-Barone-JCP2004-120-3059,spectro-Barone-JCP2005-122-014108} as implemented in Gaussian\cite{prog-G09}, we have computed the anharmonic corrections for all molecules. The largest contribution to any of the reaction enthalpies was only \SI{5}{\kJ\per\mol}, even at temperatures as high as \SI{3000}{\kelvin}, which is considered negligible for the accuracy goal of the present work. 

\subsubsection{Basis Sets and Extrapolation to the Complete Basis Set Limit}
To reach high accuracy for \textit{ab initio} thermochemistry, the computed electronic energies were extrapolated to the complete basis set limit (CBS), from two calculations with all-electron atomic natural orbitals relativistic core correlation (ANO-RCC) basis sets \cite{basis-Roos-JPC2004-108-2851,basis-Roos-CPL2005-409-295} with quality sequentially increased from triple-$\zeta$ (n=3) to quadruple-$\zeta$ (n=4). The CASSCF/HF energies were extrapolated with the Karton and Martin formula \cite{basis-Karton-TCA2006-115-330}:
\begin{align}
\label{eq:CBS_1}
E^\text{CAS/HF}_\text{n} = E^{\text{CAS/HF}}_\text{CBS} + A(n+1)\exp(-6.57\sqrt{n}),
\end{align}
separately from the correlation energies $E^\text{corr}_\text{n}$ via \cite{basis-Feller-JCP2008-129-204105,basis-Feller-JCP2011-135-044102}
\begin{align}
\label{eq:CBS_2}
E^\text{corr}_\text{n} = E^\text{corr}_\text{CBS} + \frac{B} {(n + 1/2)^{4}}.
\end{align}
While irregular basis convergence patterns was reported for lanthanide ANO-RCC basis sets~\cite{basis-Lu-JCP2016-145-054111}, the convergence is smooth in the case of the Pu basis sets as illustrated by Figure~S1 in the Supporting Information.

\subsubsection{Single Reference Wave Function Correlated Calculations}

To account for dynamic electronic correlation, in cases of open-shell systems with a weak multireference character, UCCSD(T) (unrestricted coupled cluster singles doubles and perturbative triples) \cite{ci-Watts-JCP1993-98-8718} is the gold standard. These calculations were performed using the Molpro Quantum Chemistry software \cite{molpro2015}. The frozen orbitals were the 1s orbital of O and 1s to 5d (included) orbitals of Pu. Core-valence correlation effects were discarded not only due to their high computational cost but also because we expect them to be no larger than \SI{4}{\kJ\per\mol} for actinide molecules~\cite{actinide-Bross-JCP2014-141-244308,actinide-Feng-JPC2017-121-1041}. Due to the important multiconfigurational character of the ground-state wave function of \ce{PuO3} and \ce{PuO2(OH)2} species, no UCCSD(T) calculations were performed. 

\subsubsection{Multireference Calculations and Definition of the Orbital Active Spaces}

\begin{table*}[htp]
\caption{Active Spaces for the Various Plutonium Oxides and Oxyhydroxides}
\label{Tab:active_space}
\begin{tabular}{lll}
\hline
 molecule & active space\\
 & (electron, orbitals) & orbitals\\
   \hline
  \ce{PuO2} & (4,4)\textsuperscript{\emph{a}}&  $\delta_u$(2), $\phi_u$(2)  \\
  \ce{PuO2} & (12,17)\textsuperscript{\emph{b}}& $\pi_u$(2), $\pi_g$(2), $\sigma_g$, $\sigma_u$, $\pi_u^*$(2), $\pi_g^*$(2), $\sigma_g^*$, $\sigma_u^*$, $\delta_u$(2), $\phi_u$(2), $\delta_g$(2)  \\
  \ce{PuO3}& (14,13)\textsuperscript{\emph{c}} 	& 1a$_{1}$, 2a$_{1}$/$\pi_u$, 2a$_{1}$, 4a$_{1}$/p$_{x}^{o}$, 5a$_{1}$/$\phi_{u}$, \\
  			&& 6a$_{1}$/$\pi_u$, 1b$_{1}$/$\phi_{u}$, 2b$_{1}$/$\pi_{u}$, 1b$_{2}$/p$_{z}^{o}$, 2b$_{2}$/$\sigma_u$, 3b$_{2}$/$\delta_{u}$, 4b$_{2}$/$\sigma_u^{*}$, 1a$_{2}$/$\delta_{u}$ \\
  \ce{PuO2(OH)2} & (4,4) & 1a/$\delta$, 2a/$\phi$, 1b/$\delta$, 2b/$\phi$\\
  Pu & (8,13) & 7s, 5f, 6d  \\
  O & (6,4) & 2s, 2p$_{x}$, 2p$_{y}$, 2p$_{z}$  \\
  \ce{O2} & (12, 8) & $\sigma_{2s}$, $\sigma_{2s}^{*}$, $\sigma_{2p}$, $\sigma_{2p}^{*}$, $\pi_{2p}$(2), $\pi_{2p}^*$(2)\\
  \ce{OH} & (7,5) & $\sigma$, $\sigma_{nb}$, 2p$_{x}^{O}$, 2p$_{y}^{O}$, $\sigma^{*}$,  \\
  \ce{H2O2}	& (14, 10) & 1 $\sigma^{O-O}$, 1 $\sigma^{*O-O}$, $\sigma^{O-H}$ (2), \\
                   	&& 2 $\sigma^{O-O}$, $\pi^{O-O}$, 2p$^{O}$, 1$\sigma^{*O-O}$, $\sigma^{O-H}$(2)  \\
  \ce{H2O} & (8,6) & 2a$_{1}$,1b$_{2}$, 3a$_{1}$, 1b$_{1}$, 4a$_{1}$, 1b$_{2}$ \\
  \ce{H2} & (2, 2) & $\mathrm{\sigma_g}$, $\mathrm{\sigma_u^{*}}$\\
  H & (1,1) & 1s\\
\hline
\end{tabular}
\textsuperscript{\emph{a}} For the ground-state energy calculation.\\
\textsuperscript{\emph{b}} For ground and excited-states that are coupled by the spin-orbit Hamiltonian in the RASSI.\\
\textsuperscript{\emph{c}} The orbital numbering refers to the active space, e.g. 1a$_1$ is the first active orbital of a$_1$ symmetry.
\end{table*}

Knowing the multi-configurational character of the ground-state wave function of the atomic plutonium and the plutonium oxides and oxyhydroxides of interest \cite{actinide-Boguslawski-PCCP2017-19-4317}, and since SOC needs to be included \textit{a posteriori} \cite{actinide-Gagliardi-CSR2007-36-893}, the use of a multi-configurational approach that accounts for the static and dynamic electronic correlation effects is mandatory. The former was included with the CASSCF (Complete Active Space Self Consistent Field) method \cite{mcscf-Knowles-CPL1985-115-259,mcscf-Werner-JCP1985-82-5053}. The active spaces are listed in the Table~\ref{Tab:active_space}; for the atoms (O, H, Pu) and molecules without Pu (OH, \ce{H2O2}, \ce{H2}, \ce{H2O}), the active space is composed of the valence orbitals. Nevertheless, one has to note that the CAS (Complete Active Space) of (8,13) for the plutonium atom lead to convergence problems as the competing configurations $\mathrm{5f^67s^2}$, $\mathrm{5f^66d^17s^1}$, $\mathrm{5f^66d^17s^27p^1}$ are close in energy and intertwined.\cite{actinide-Bovey-JOSA1961-51-522,actinide-Blaise-CAF1962-255-2403,actinide-Blaise-PS1980-22-224} We instead used a RAS (Restricted Active Space) \cite{mcscf-Malmqvist-JPC1990-94-5477} to focus on the two configurations $\mathrm{5f^67s^2}$, $\mathrm{5f^66d^17s^1}$ that dominate the low-lying part of the Pu spectrum (up to \SI{18000}{\per\cm}. The RAS space includes the 7s in the RAS1, the $\mathrm{5f}$ in the RAS2 and the 6d in RAS3 with one as the maximum number of holes in RAS1 and as number of particles in RAS3.

For the plutonium molecules, the difficulty is to design orbital active spaces that do not exceed the current active space limit of about 16 electrons in 16 orbitals. In a recent work, some of us have used the multireference density matrix renormalization group (DMRG) calculations \cite{actinide-Boguslawski-PCCP2017-19-4317}, in which we could afford including all the valence orbitals from both Pu and oxygen atoms into the active space, thus allowing the maximum flexibility into the ground-state wave functions. These DMRG calculations cannot be used for quantitative thermodynamics calculations, as they were restricted to a DMRG-CI approach, and did not include orbital relaxation, dynamic correlation and SOC. However, concepts of quantum information theory, such as orbital-pair correlations, were helpful to identify which orbitals contribute most to the ground-state wave functions, helping the design of reduced ``optimal'' active spaces \cite{actinide-Boguslawski-PCCP2017-19-4317} that can be reasonably handled in the SO-CASPT2 calculations carried out here . 

Regarding \ce{PuO2} (symmetry $D_{2h}$), the choice of the active space to describe the low lying excited states was based on the work of Denning~{\etal} \cite{actinide-Denning-JPC2007-111-4125} and La Macchia~{\etal} \cite{actinide-La-Macchia-PCCP2008-10-7278}. The full valence active space is composed of 2 $\pi_u$, 2 $\pi_g$, $\sigma_g$, and $\sigma_u$ bonding orbitals, which are combinations of 5f and 6d orbitals of plutonium mixed with the 2p orbitals of the oxygen atoms, the corresponding antibonding orbitals, and the four nonbonding $\delta_u$, $\phi_u$. 

Despite the increase of the computational power, a CASSCF calculation with the full valence active space \cite{actinide-Boguslawski-PCCP2017-19-4317} made of 16 electrons in 19 orbitals is still not feasible (\num{203 440 360}~CSF for the $^5\Sigma_g^+$ ground state). We thus chose to remove two lowest doubly occupied $\pi_g$ orbitals, leading to an active space of 12 electrons in 17 orbitals. Note that La Macchia~{\etal} \cite{actinide-La-Macchia-PCCP2008-10-7278} used a smaller active space of 12 electrons in 14 orbitals, as they discarded the $\pi_g^*$ and $\sigma_g^*$ antibonding orbitals. Whereas for a good estimate of SOC effect in a two-step approach, numerous excited states have to be computed (to position accurately the excited states with respect to the ground state), this could degrade the description of the ground state in terms of total energy and wave function \cite{mcscf-Starling-MP2001-99-103}. Thus the ground-state $\mathrm{^5\Sigma_g^+}$ spin-free energy is computed from the well-suited minimal active space that distributes four electrons in the four nonbonding orbitals ($\delta_u$ and $\phi_u$), denoted SO-CASPT2(4,4).

To define the important orbitals in the active space of \ce{PuO3} (symmetry $C_{2v}$), we used information from \ce{PuO2^{2+}} that can be considered as a subunit of plutonium trioxide suggesting an active space composed of two $\pi$ orbitals, one $\sigma$ orbital along with the corresponding antibonding associated to the plutonyl \ce{PuO2^{2+}} subunit, the four nonbonding  $\delta_u$ and $\phi_u$ orbitals and the $\mathrm{p_x}$, $\mathrm{p_y}$ and $\mathrm{p_z}$ of the third distant oxygen; this sums up to an active space of 14 electrons distributed in 13 orbitals as depicted in ref. \citenum{actinide-Boguslawski-PCCP2017-19-4317}. The active space to treat the cationic form \ce{PuO3+} includes one electron less.

For \ce{PuO2(OH)2} (symmetry $C_2$), the strengths of the orbital-pair correlations depicted in ref. \citenum{actinide-Boguslawski-PCCP2017-19-4317} indicate that the nonbonding $\delta_u$ and $\phi_u$ orbitals of Pu are the most entangled orbitals and should be part of a chemically relevant active space for the ground-state wave function as well as the low-lying excited states. This yielded 16 electronic excited spin-free states. For the other Pu systems, it was not possible to include all states. Indeed, the higher excited states turned out to have too low reference weights (<20\%) in the Multi-State Complete Active Space with Second-Order Perturbation Theory (MS-CASPT2) \cite{MS-CASPT2} wave-functions.

The case of the Pu atom was peculiar since the RASSCF was not able to reproduce the exact degeneracy of the states of the same $L$ value. Therefore, the degeneracy was manually imposed across states with identical $L$ and $S$ values. The CASSCF and MS-CASPT2 calculations finally include 80, 119, and 16 electronic states for \ce{PuO2}, \ce{PuO3}, and \ce{PuO2(OH)2}, respectively as detailed in Table~S6 in the Supporting Information. As for the UCCSD(T) calculations the 1s orbital of O and the 1s to 5d orbitals (included) for Pu were kept frozen. To guarantee a good description of all the states, an imaginary shift of \SI{0.05}{\astronomicalunit} \cite{imaginary} was added to the zeroth-order Hamiltonian in addition to the IPEA = \SI{0.25}{\astronomicalunit} correction \cite{mrpt2-Ghigo-C2004-396-142}.
 
\subsubsection{Spin-Orbit Coupling State Interaction Method} 

For the Pu atom and the plutonium molecules, SOC was treated \textit{a posteriori} by state interactions between the MS-CASPT2 wave functions, using the restricted active space state interaction (RASSI) program \cite{RASSI-Malmqvist-CPL2002-357-230}, with spin-orbit integrals computed in the atomic mean field approximation using the AMFI code \cite{so-Hess-CPL1996-251-365-371,prog-schimmelpfennig-amfi1996}. These results are referred to as SO-CASPT2. The SOC contributions to the ground-state energies, i.e. the difference between the ground-state spin-free MS-CASPT2 energies (SF-CASPT2) and the SO-CASPT2 energies, was added to all other scalar relativistic results, such as the DFT and UCCSD(T) results. In Table~\ref{Tab:PuOx_Hform}, $\Delta E_{SO}$ corresponds to the difference between the SOC contributions of the plutonium molecule and the atomic Pu. All SO-CASPT2 calculations were performed with the MOLCAS8 package \cite{molcas8}.

\subsection{One-step relativistic correlated calculations}
\label{2c-details}

We also performed one-step relativistic calculations based on the molecular mean-field \cite{Sikkema2009} approximation to the Dirac--Coulomb ({\dcm}) Hamiltonian, at MP2, CCSD and CCSD(T) \cite{CCSD-Visscher-JCP1996-105-8769-8776,correlation-Visscher-JCP2001-115-9720} levels of theory with \textsc{Dirac18} electronic structure code \cite{DIRAC18} for \ce{Pu}, \ce{PuO2}, \ce{H2}, \ce{OH}, \ce{H2O} and \ce{H2O2}. This allowed us to obtain the energies for reactions R$_3$--R$_6$ yielding \ce{PuO2}. 

The Dyall basis sets \cite{basis-Dyall-TCA2012-131-3962,basis-Dyall-TCA2007-117-491} of triple- and quadruple-zeta quality were employed for the plutonium atom and Dunning aug-cc-pVnZ ($n=3, 4$) sets \cite{basis-Dunning-JCP1989-90-1007} for the light elements, all of which are left uncontracted. These calculations have been performed in $D_{\infty h}$ symmetry for \ce{Pu}, \ce{PuO2} and \ce{H2}, in $C_{\infty v}$ symmetry for \ce{OH}, in $C_{2v}$ symmetry for \ce{H2O} and in $C_1$ symmetry for \ce{H2O2}. In all cases, we've employed the approximation of the $\left(SS|SS \right)$-type two-electron integrals by a point-charge model \cite{relat-Visscher-TCA1997-98-68}. Electrons in molecular spinors with energies between -3 and \SI{100}{\astronomicalunit} were included in the correlated treatment, which amounted to correlating 28 electrons for \ce{PuO2}, 16 for \ce{Pu}, 12 for \ce{H2O2}, 6 for \ce{OH} and \ce{H2O}, and 2 for \ce{H2}. 

For \ce{PuO2}, the SCF calculations were closed-shell ones, with 52 and 58 electrons in g and u irreducible representations, respectively. For this species, the equilibrium structure and stretching frequencies were determined via a polynomial fit of the potential energy surface constructed from single-point calculations for symmetric and antisymmetric \ce{Pu-O} stretching modes. Because of constraints in computational resources, the \ce{O-Pu-O} bending mode frequency was obtained from {\dcm}-CCSD(T) calculations correlating a smaller number of spinors between -3 and \SI{5}{\astronomicalunit}. Energies and potential energy curves at the complete basis set limit were obtained with the same formula (Eqs.~\ref{eq:CBS_1} and \ref{eq:CBS_2}) introduced previously.

The Pu atom is converged at the Hartree-Fock level to the $J=0$ ground-state, with 44 and 50 electrons in the g and u irreducible representations, respectively. For the hydroxy radical, the SCF  calculation was performed using the average-of-configurations formalism \cite{relat-Thyssen-PhD2001}, with 8 electrons taken as closed shell and 1 as open shell. In the open-shell coupled cluster model, the occupations of the different irreducible representations ($\Omega = 1/2,-1/2,3/2,-3/2$) were restricted to be 3, 3, 1, 0, respectively. The internuclear distance used was that obtained at the DFT level (see section~\ref{comp:geometries}), as was the case for \ce{H2}, \ce{H2O}, and \ce{H2O2}.


Apart from the ground-state MP2, CCSD, and CCSD(T) calculations, we performed equation-of-motion calculations for ionization potentials (EOM-IP-CCSD) for atomic \ce{Pu} and \ce{PuO2}, as well as excitation energies (EOM-EE-CCSD) \cite{CCmethods-Shee-JCP2018-147-174113} for \ce{PuO2}, on the basis of the coupled cluster wave functions above.

In the case of EOM-IP-CCSD for \ce{Pu}, we requested the number of states as follows: five $\mathrm{\Omega = {1/2}_g}$, two $\mathrm{\Omega = {3/2}_g}$, one $\mathrm{\Omega ={5/2}_g}$, three $\mathrm{\Omega ={1/2}_u}$, two $\mathrm{\Omega ={3/2}_u}$ and one $\mathrm{\Omega ={5/2}_u}$, whereas for \ce{PuO2} we requested two $\mathrm{\Omega = {1/2}_g}$, one $\mathrm{\Omega = {3/2}_g}$, three $\mathrm{\Omega ={1/2}_u}$, two $\mathrm{\Omega ={3/2}_u}$ and one $\mathrm{\Omega ={5/2}_u}$ states. In the case of EOM-EE-CCSD we requested  the number of states as follows: six $\mathrm{\Omega = {0}_g}$, four $\mathrm{\Omega = {1}_g}$, three $\mathrm{\Omega ={2}_g}$, two $\mathrm{\Omega ={3}_g}$, one $\mathrm{\Omega ={4}_g}$, one $\mathrm{\Omega ={5}_g}$,  ten $\mathrm{\Omega = {0}_u}$, nine $\mathrm{\Omega = {1}_u}$, seven $\mathrm{\Omega ={2}_u}$, five $\mathrm{\Omega ={3}_u}$, three $\mathrm{\Omega ={4}_u}$, and one $\mathrm{\Omega ={5}_u}$.

\section{Results and Discussion}\label{Sec:Results_and_discussion}

\subsection{Electronic Structure Analysis}
\label{Sec:electronic_structure}
\subsubsection {Atomic \ce{Pu} }

\begin{table*}[htp]
\caption{Fine Structure Transition Energies ($\Delta E$ in \si{\per\cm}) of Atomic Pu Computed at the SO-CASPT2 with the ANO-RCC-TZVP Basis Set and Analysis of the Various $J$-States in Terms of the Dominating $LS$ Terms}
\label{Tab:levelsPu}
\begin{tabular}{*2S*2l}
\hline
  {exptl \cite{actinide-Bovey-JOSA1961-51-522,actinide-Blaise-CAF1962-255-2403,actinide-Blaise-PS1980-22-224}} & {$\Delta E$}    & $J$-value &  weight of $LS$ states	\\
  \hline
0 & 0   &  0  &  54\%   $\mathrm{^{7}F}$,  38\%   $\mathrm{^{5}D}$,  7\%   $\mathrm{^{3}P}$ \\
2 204 &1806  &  1  &  64\%   $\mathrm{^{7}F}$,  32\%   $\mathrm{^{5}D}$                                      \\
4 300 &  4208  &  2  &  82\%   $\mathrm{^{7}F}$,  17\%   $\mathrm{^{5}D}$                                            \\
6 145 &6353  &  3  &  88\%   $\mathrm{^{7}F}$,  10\%   $\mathrm{^{5}D}$                                            \\
7 775 &8091  &  4  &  93\%   $\mathrm{^{7}F}$                                                  \\
9 179 & 9552  &  5  &  90\%   $\mathrm{^{7}F}$                                            \\
10 238  & 11010  &  6  &  85\%   $\mathrm{^{7}F}$,  10\%   $\mathrm{^{5}G}$                                            \\                      
9 773 & 12499  &  0  &  44\%   $\mathrm{^{7}F}$,    21\%   $\mathrm{^{5}D}$,  8\%   $\mathrm{^{3}P}$                                \\
13 528 &13612  &  1  &  66\%   $\mathrm{^{9}H}$, 9\%   $\mathrm{^{7}F}$,      7\%   $\mathrm{^{7}G}$                                \\
\hline                                                
\end{tabular}
\end{table*} 
To compute the enthalpies of formation of the plutonium molecules, an accurate description of the plutonium atom, i.e. its electronic structure, is mandatory. Thus, in Table~\ref{Tab:levelsPu} and Table~S2 in the Supporting Information, the SO-CASPT2 calculations are reported and compared to available data from the literature \cite{actinide-Bovey-JOSA1961-51-522,actinide-Blaise-CAF1962-255-2403,actinide-Blaise-PS1980-22-224}. The SO-CASPT2 energy levels are able to accurately reproduce the experimental assignments with errors between 91 and \SI{397}{\per\cm} for the first five excited states below \SI{9500}{\per\cm}. Deviations between our results and the available data appear after the sixth state; the experimental assignment predicts the sixth state at \SI{9773}{\per\cm} with a $J=0$ while such state is found at \SI{12488}{\per\cm} in the current study though uncertainties remain for the attribution of this electronic state. The computed seventh state with $J=6$ is about \SI{800}{\per\cm} higher energy than the experimental value. Looking closely at the nature of the plutonium low-lying states, one can first notice the remarkable mixing between the $\mathrm{^{7}F}$, $\mathrm{^{5}D}$ and $\mathrm{^{3}P}$ spin-orbit free states and that the contribution of the $\mathrm{^{5}D}$ state decreases as the energy increases.  Finally, the most important result is the overall good reproduction of the SO splitting of $\mathrm{^{7}F}$ spin-free state, making us confident about the accuracy of the SOC for the plutonium molecules. We further note that we have the single-reference $J=0$ ground state, well described by the {\dcm}-CCSD(T) method.

\subsubsection {\ce{PuO3} and \ce{PuO2(OH)2} molecules}

\begin{figure}[htp]
\includegraphics[width=\linewidth]{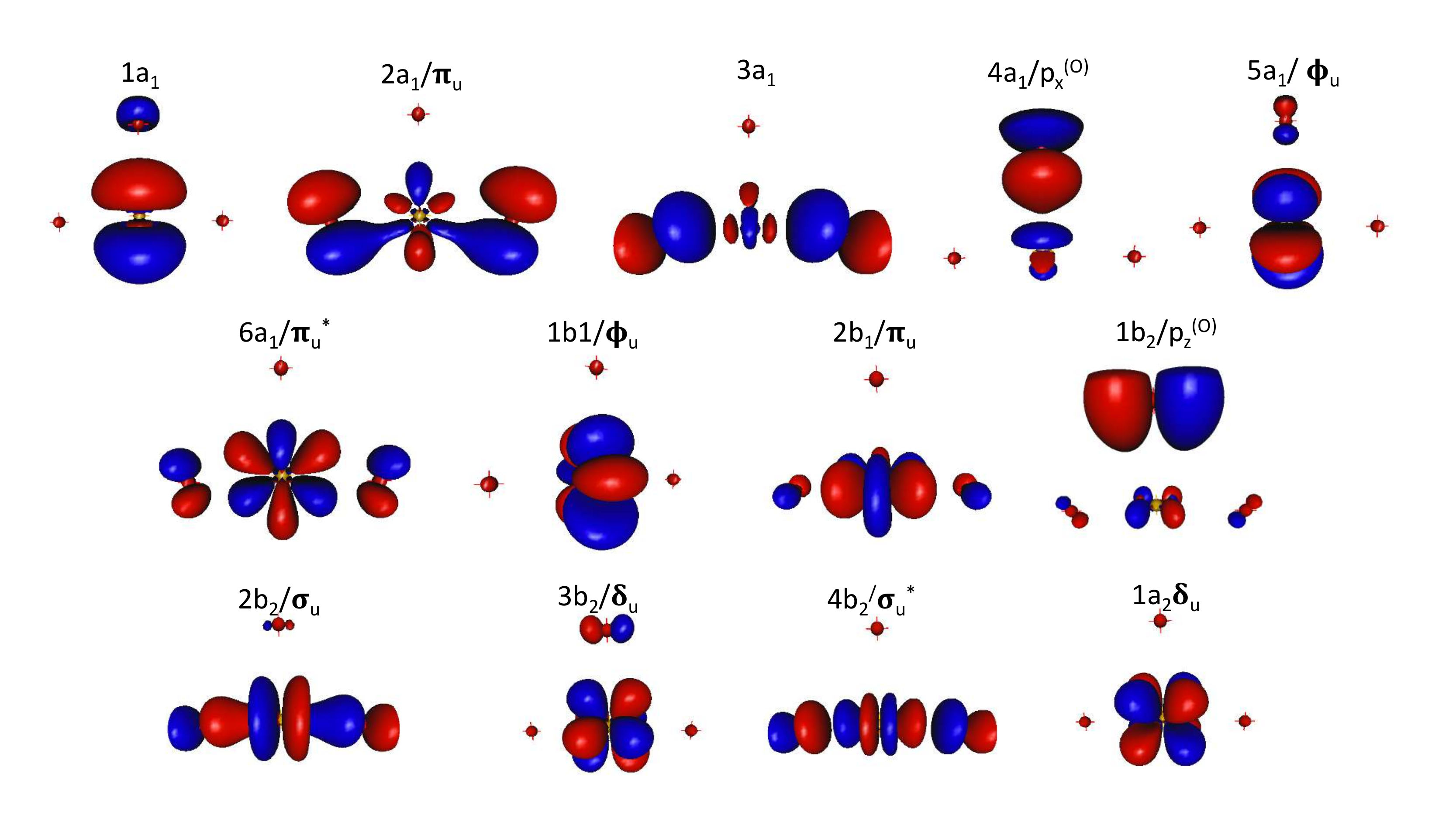}
\caption{Active CASSCF orbitals of \ce{PuO3} at the CASSCF level; Isosurface = \SI{0.05}{\astronomicalunit}; Distances: Pu-O(PuO$_2^{2+}$subunit) = \SI{1.767}{\angstrom}; Pu-O$_{axial}$ = \SI{1.934}{\angstrom}; angle (O-Pu-O$_{axial}$) = 93.86\textdegree}
\label{Fig:MolOrbPuO3}
\end{figure}

\begin{table*}[htp]
\begin{footnotesize}
\caption{Lowest Vertical Transition Energies ($\Delta E$ in \si{\per\cm}) of \ce{PuO3} Computed at the SF-CASPT2 and SO-CASPT2 Levels with the ANO-RCC-TZVP Basis Set and Analysis of the Various States}
\label{Tab:levelsPuO3}
\begin{tabular}{lllSl}
\hline
 & no.  &  state  &  {$\Delta E$}  &  character  : \% [orbital(number of electron)]\\
\hline
SF &1  &  $\mathrm{^3B_2(1)}$  &  0  &  36\% [4a$_1$/p$^{(O)}$(2), 1b$_1$/$\mathrm{\phi_u(1)}$, 1b$_2$/p$^{(O)}$($\overline{1}$), 2b$_2$/$\mathrm{\sigma_u(2)}$, 3b$_2$/$\mathrm{\delta_u(1)}$, 1a$_2$/$\mathrm{\delta_u(1)}$]\\
& & & & 21\% [4a$_1$/p$^{(O)}$(2), 1b$_1$/$\mathrm{\phi_u(1)}$, 1b$_2$/p$^{(O)}$(1), 2b$_2$/$\mathrm{\sigma_u(2)}$, 3b$_2$/$\delta_{u}$($\overline{1}$), 1a$_2$/$\mathrm{\delta_u(1)}$]\\
 & & & & 22\% [4a$_1$/p$^{(O)}$(2), 1b$_1$/$\mathrm{\phi_u(1)}$, 1b$_2$/p$^{(O)}$(2), 2b$_2$/$\mathrm{\sigma_u(2)}$, 1a$_2$/$\mathrm{\delta_u(1)}$]\\
& 2  &  $\mathrm{^3A_2(1)}$  &  3161  & 43\% [4a$_1$/p$^{(O)}$(2), 5a$_1$/$\mathrm{\phi_u(1)}$, 1b$_2$/p$^{(O)}$($\overline{1}$), 2b$_2$/$\mathrm{\sigma_u(2)}$,3b$_2$/$\mathrm{\delta_u(1)}$, 1a$_2$/$\mathrm{\delta_u(1)}$]\\
& & & & 23\% [4a$_1$/p$^{(O)}$(2), 5a$_1$/$\mathrm{\phi_u(1)}$, 1b$_2$/p$^{(O)}$(1), 2b$_2$/$\mathrm{\sigma_u(2)}$,3b$_2$/$\delta_{u}$($\overline{1}$), 1a$_2$/$\mathrm{\delta_u(1)}$]\\
& & & & 17\% [4a$_1$/p$^{(O)}$(2), 5a$_1$/$\mathrm{\phi_u(1)}$, 1b$_2$/p$^{(O)}$(2), 2b$_2$/$\mathrm{\sigma_u(2)}$, 1a$_2$/$\mathrm{\delta_u(1)}$]\\
& 3  &  $\mathrm{^3B_1(1)}$  &  4663  &  38\% [4a$_1$/p$^{(O)}$(2),5a$_1$/$\mathrm{\phi_u(1)}$, 1b$_1$/$\mathrm{\phi_u(1)}$, 1b$_2$/p$^{(O)}$($\overline{1}$), 2b$_2$/$\mathrm{\sigma_u(2)}$,3b$_2$/$\mathrm{\delta_u(1)}$] \\
& & & & 12\% [4a$_1$/p$^{(O)}$(2), 5a$_1$/$\mathrm{\phi_u(1)}$, 2b$_1$/$\pi_{u}$(1), 1b$_2$/p$^{(O)}$($\overline{1}$), 2b$_2$/$\mathrm{\sigma_u(2)}$,3b$_2$/$\mathrm{\delta_u(1)}$] \\
& & & & 12\% [4a$_1$/p$^{(O)}$(2), 5a$_1$/$\mathrm{\phi_u(1)}$, 1b$_1$/$\mathrm{\phi_u(1)}$, 2b$_1$/$\pi_{u}$(2), 1b$_2$/p$^{(O)}$(2)] \\
& 4  &  $\mathrm{^1A_1(1)}$  &  7239  & 32\% [4a$_1$/p$^{(O)}$(2), 1b$_2$/p$^{(O)}$(1), 2b$_2$/$\mathrm{\sigma_u(2)}$,3b$_2$/$\delta_{u}$($\overline{1}$), 1a$_2$/$\mathrm{\delta_u(2)}$]\\
&&  & & 28\% [4a$_1$/p$^{(O)}$(2), 1b$_2$/p$^{(O)}$(2), 2b$_2$/$\mathrm{\sigma_u(2)}$, 1a$_2$/$\mathrm{\delta_u(2)}$]\\
& & & & 11\% [4a$_1$/p$^{(O)}$(1), 5a$_1$/$\Phi_{u}$($\overline{1}$), 1b$_2$/p$^{(O)}$(2), 2b$_2$/$\mathrm{\sigma_u(2)}$, 1a$_2$/$\mathrm{\delta_u(2)}$]\\
& 5  &  $\mathrm{^3A_2(2)}$  &  7287  &  20\%  [4a$_1$/p$^{(O)}$(2), 1b$_1$/$\mathrm{\phi_u(1)}$, 1b$_2$/p$^{(O)}$(1), 2b$_2$/$\mathrm{\sigma_u(2)}$,3b$_2$/$\mathrm{\delta_u(2)}$] \\
& & & & 18\%  [4a$_1$/p$^{(O)}$(1),5a$_1$/$\Phi_{u}$($\overline{1}$), 1b$_1$/$\mathrm{\phi_u(1)}$, 1b$_2$/p$^{(O)}$(2), 2b$_2$/$\mathrm{\sigma_u(2)}$,3b$_2$/$\mathrm{\delta_u(1)}$] \\
& & & & 17\%  [4a$_1$/p$^{(O)}$(2), 1b$_1$/$\mathrm{\phi_u(1)}$, 1b$_2$/p$^{(O)}$(2), 2b$_2$/$\mathrm{\sigma_u(2)}$,3b$_2$/$\mathrm{\delta_u(1)}$] \\
& 6  &  $\mathrm{^3B_1(2)}$  &  8038  &  31\% [4a$_1$/p$^{(O)}$(1),5a$_1$/$\Phi_{u}$($\overline{1}$), 1b$_2$/p$^{(O)}$(2), 2b$_2$/$\mathrm{\sigma_u(2)}$,3b$_2$/$\mathrm{\delta_u(1)}$, 1a$_2$/$\mathrm{\delta_u(1)}$] \\
& & & & 11\% [4a$_1$/p$^{(O)}$(2),5a$_1$/$\mathrm{\phi_u(1)}$, 2b$_1$/$\pi_{u}$(1), 1b$_2$/p$^{(O)}$($\overline{1}$), 2b$_2$/$\mathrm{\sigma_u(2)}$,3b$_2$/$\mathrm{\delta_u(1)}$]\\
& & & & 10\% [4a$_1$/p$^{(O)}$(2), 1b$_2$/p$^{(O)}$(2), 2b$_2$/$\mathrm{\sigma_u(2)}$,3b$_2$/$\mathrm{\delta_u(1)}$, 1a$_2$/$\mathrm{\delta_u(1)}$]\\
& & & & 10\% [4a$_1$/p$^{(O)}$(2), 1b$_1$/$\mathrm{\phi_u(1)}$, 2b$_1$/$\pi_{u}$(1), 1b$_2$/p$^{(O)}$($\overline{1}$), 2b$_2$/$\mathrm{\sigma_u(2)}$, 1a$_2$/$\mathrm{\delta_u(1)}$]\\
\hline
SO &  & X  & 0  &  47\%  $\mathrm{^3B_2(1)}$  +  24\%  $\mathrm{^3A_2(1)}$  +  14\%  $\mathrm{^3B_1(1)}$    \\
&   &a  &1235  &  59\%  $\mathrm{^3B_2(1)}$  +  25\%  $\mathrm{^3A_2(1)}$                    \\
&   &b &  1783  &  73\%  $\mathrm{^3B_2(1)}$                          \\
&   &c  &3777  &  51\%  $\mathrm{^3A_2(1)}$                          \\
&   &d  &5660  &  24\%  $\mathrm{^3B_1(1)}$  + 18\%  $\mathrm{^1A_1(1)}$  + 17\%  $\mathrm{^3A_2(2)}$  +  10\%  $\mathrm{^3B_1(2)}$  \\  
\hline
\end{tabular}
\end{footnotesize}
\end{table*}

For \ce{PuO3}, at the spin-orbit level, we report the description of the ground state and the six lowest excited states in Table~\ref{Tab:levelsPuO3}. It is noteworthy that the $\mathrm{1a_g}$,  $\mathrm{2a_g}$/$\mathrm{\delta_u}$ and 3a$_{g}$ orbitals are doubly occupied in all the calculated spin-free states. The \ce{PuO3} orbitals are composed of a mixture of orbitals belonging to the plutonyl subunit and the distant oxygen. Thus, for the sake of clarity, we label these orbitals with those normally associated to the \ce{PuO2^2+} ion (with linear energy $D_{2h}$ notations), and those of the distant oxygen atom are denoted by p$_{x,y,z}^{(O)}$.  The SO ground state is composed of 47\% $\mathrm{^3B_2}$, 24\% $\mathrm{^3A_2}$  and 14\% $\mathrm{^3B_1}$ spin-free states. The composition is similar to the one previously reported by Kovács \cite{actinide-Kovacs-JPC2017-121-2523} (52\% $\mathrm{^3B_2}$ and 24\% $\mathrm{^3A_2}$). More striking is the difference between our computed vertical excitation energies with that of Kovács's SO-CASPT2 spectrum \cite{actinide-Kovacs-JPC2017-121-2523}. 

We predict the first excited state to be at \SI{1235}{\per\cm}, in contrast to the \SI{471}{\per\cm} previously found \cite{actinide-Kovacs-JPC2017-121-2523}, and the overall spectrum determined from our calculations is much denser and lower by about \SI{4000}{\per\cm} than in ref. \citenum{actinide-Kovacs-JPC2017-121-2523}. Such discrepancy may find its origin in the description of the spin-free states. Although we agree with Kovács that the spin-free ground state has $\mathrm{^3B_2}$ symmetry, we find a different orbital character. In our calculations, the $\mathrm{^3B_2}$ exhibits a strong multideterminantal character with 57\% corresponding to the configuration in which the 4a$_1$/p$_z^{(O)}$ and 2b$_2$/$\sigma_u$ orbitals are doubly occupied (See Table~\ref{Tab:levelsPuO3} and Figure~\ref{Fig:MolOrbPuO3}) and the singly occupied orbitals 5b$_1$/$\Phi_u$, 1b$_2$/p$_z^{(O)}$, 3b$_2$/$\delta_u$ and 1a$_2$/$\delta_u$ and 22\% from the electronic configuration in which of the 4a$_1$/p$_z^{(O)}$, 1b$_2$/p$_z^{(O)}$ and 2b$_2$/$\sigma_u$ orbitals are doubly occupied and the 5b$_1$/$\Phi$ and 1a$_2$/$\delta_u$ singly occupied. Kovács, however, finds the ground state to be mainly composed of plutonyl orbitals (79\%). 

Such a difference can be explained by the fact that the considered active spaces are different, a CAS(10,16) in his case and a CAS(14,13) in our work. In a close look at the entanglement diagram of \ce{PuO3} derived from DMRG calculations \cite{actinide-Boguslawski-PCCP2017-19-4317} (Figure~6a, Full-Valence CAS) there is a strong correlation within the nonbonding plutonyl-like orbitals ($\delta_u, \phi_u$) and also between the b$_2$/p$^{(O)}$  and the $\mathrm{5f\delta_u}$ orbitals, in agreement with our analysis of the $\mathrm{^3B_2}$ ground-state wave function.

\begin{figure}[htp]
\includegraphics[width=\linewidth]{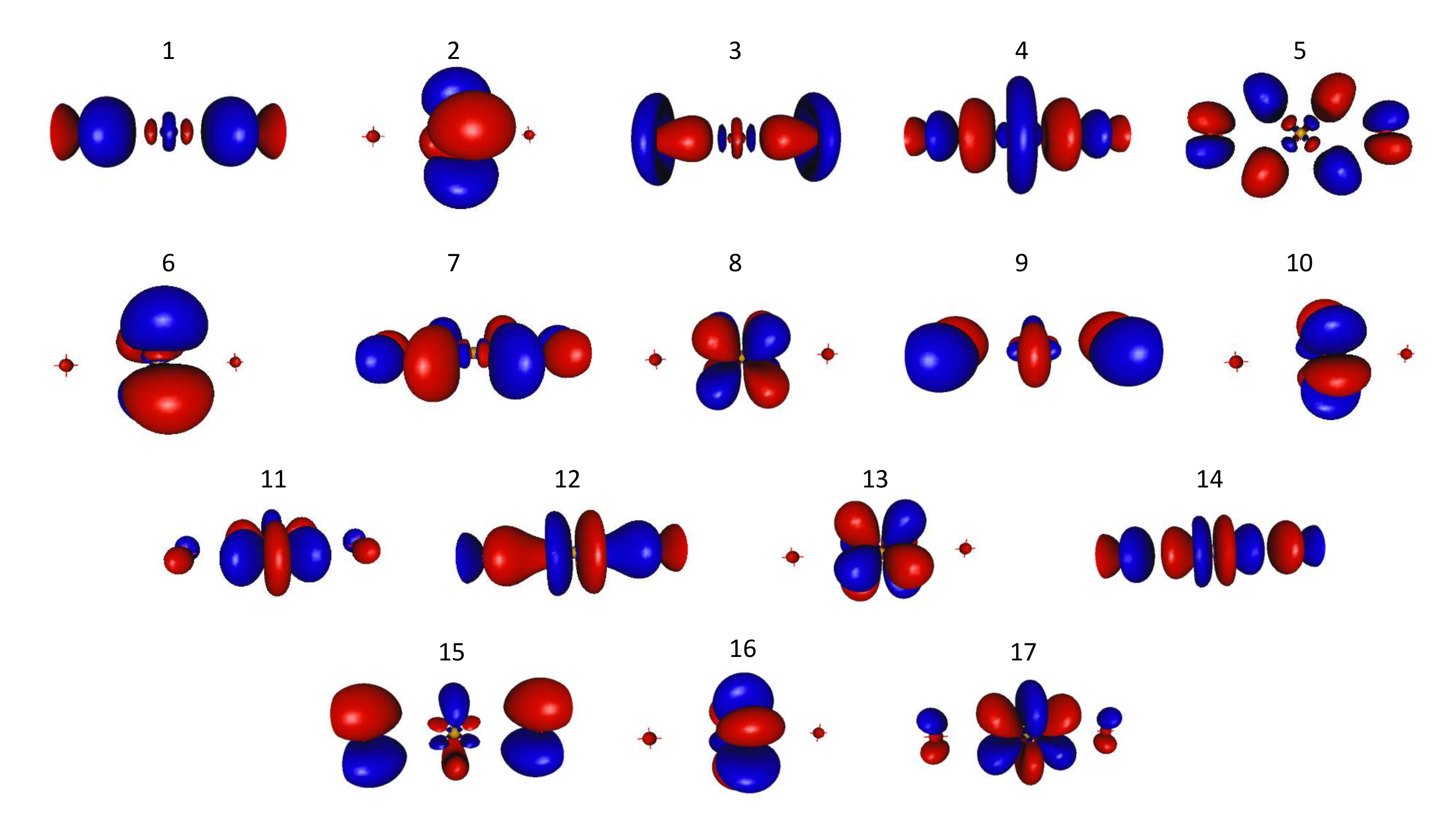}
\caption{Molecular orbitals of \ce{PuO2(OH)2} at CASSCF level. Isosurface = \SI{0.05}{\astronomicalunit}; Distances: Pu-OH = \SI{2.103}{\angstrom}, Pu-O = \SI{1.742}{\angstrom}; Angles: O-Pu-OH = \SI{92.15}{\degree}; HO-Pu-OH = \SI{105.68}{\degree}; O-Pu-O = \SI{174.29}{\degree}.}
\label{Fig:MolOrbPuO2OH2}
\end{figure}

\begin{table*}[htp]
\caption{Lowest Vertical Transition Energies ($\Delta E$ in \si{\per\cm}) of \ce{PuO2(OH)2}, Computed at the SF-CASPT2 and SO-CASPT2 Levels with ANO-RCC-TZVP Basis Sets and Analysis of the Various States}
\label{Tab:levelsPuO2OH2}
\begin{tabular}{ccSl}
\hline
  &  state  &  {$\Delta E$}   &  character  : \% [orbital(number of electron)]\\
\hline
SF&   $\mathrm{(1)^{3}B}$  &    0 &  49\% [2a/$\mathrm{\phi_u(1)}$, 1b/$\mathrm{\delta_u(1)}$]  \\
&    &    &  48\% [1a/$\mathrm{\delta_u(1)}$,   2b/$\mathrm{\phi_u(1)}$]  \\
&  $\mathrm{(1)^{3}A}$  &  889  &    41\% [1a/$\mathrm{\delta_u(1)}$,           2a/$\mathrm{\phi_u(1)}$]  \\
&    &    &    59\% [1b/$\mathrm{\delta_u(1)}$,             2b/$\mathrm{\phi_u(1)}$]  \\
&   $\mathrm{(2)^{3}B}$&  2136    &  75\% [1a/$\mathrm{\delta_u(1)}$,   1b/$\mathrm{\delta_u(1)}$]  \\
&    &    &  20\% [2a/$\mathrm{\phi_u(1)}$, 2b/$\mathrm{\phi_u(1)}$]   \\
&  $\mathrm{(2)^{3}A}$   &  4698  &    59\% [1a/$\mathrm{\delta_u(1)}$,             2a/$\mathrm{\phi_u(1)}$]  \\
&    &    &    41\%  [1b/$\mathrm{\delta_u(1)}$,           2b/$\mathrm{\phi_u(1)}$]  \\
&  $\mathrm{(3)^{3}B}$  &5351    &  52\% 2a/$\mathrm{\phi_u(1)}$, 1b/$\mathrm{\delta_u(1)}$    \\
&    &    &  46\% [1a/$\mathrm{\delta_u(1)}$,   2b/$\mathrm{\phi_u(1)}$]  \\
&  $\mathrm{(1)^{1}A}$  &5764    &  [58\%  1b/$\mathrm{\delta_u(1)}$,           2b/$\Phi_{u}$($\overline{1}$)]  \\
&    &    &  30\%  [1a/$\mathrm{\delta_u(1)}$,           2a/$\Phi_{u}$($\overline{1}$)]  \\
&    &    &  5\%[2a/$\mathrm{\phi_u(2)}$]   \\
&  $\mathrm{(1)^{1}B}$  &  7444  &  50\% [1a/$\mathrm{\delta_u(1)}$,   1b/$\delta_{u}$($\overline{1}$)]  \\
&    &    &  28\% [2a/$\mathrm{\phi_u(1)}$, 2b/$\Phi_{u}$($\overline{1}$)]    \\
&    &    &  16\%[ 1a/$\mathrm{\delta_u(1)}$,   2b/$\Phi_{u}$($\overline{1}$)]  \\
&$\mathrm{(2)^{1}A}$  &    9626  &  39\%  [1a/$\mathrm{\delta_u(1)}$,           2a/$\Phi_{u}$($\overline{1}$)]  \\
&    &    &  29 \%   [2a/$\mathrm{\phi_u(2)}$]  \\
&    &    &  13\% [1b/$\mathrm{\delta_u(1)}$,           2b/$\Phi_{u}$($\overline{1}$)]  \\
&    &    &  10\% [1a/$\mathrm{\delta_u(2)}$]  \\
&    &    &  7 \%[2a/$\mathrm{\phi_u(2)}$]   \\
\hline                       
SO  & X   &  0  &  46\%  $\mathrm{(1)^{3}A}$,  41\%  $\mathrm{(1)^{3}B}$,  6\%  $\mathrm{(2)^{3}B}$              \\
& a &  323  &  45.\%  $\mathrm{(1)^{3}B}$,  45\%  $\mathrm{(1)^{3}A}$,  6\%  $\mathrm{(2)^{3}B}$              \\
& b &   2222  &  35\%  $\mathrm{(2)^{3}B}$,  25\%  $\mathrm{(1)^{3}B}$,  14\%  $\mathrm{(1)^{1}A}$,  13\%  $\mathrm{(2)^{3}A}$,  10\%  $\mathrm{(3)^{3}B}$  \\
& c &  3757  &  40\%  $\mathrm{(2)^{3}A}$,  25\%  $\mathrm{(2)^{3}B}$,  19\%  $\mathrm{(1)^{1}B}$,  7\%  $\mathrm{(1)^{3}B}$,  6\%  $\mathrm{(1)^{3}A}$  \\
& d &  4158  &  36\%  $\mathrm{(3)^{3}B}$,  34\%  $\mathrm{(2)^{3}B}$,  14\%  $\mathrm{(2)^{1}A}$,  11\%  $\mathrm{(1)^{3}B}$,      \\
\hline
\end{tabular}
\end{table*}

Concerning \ce{PuO2(OH)2}, the low-energy part of the vertical spectrum including SOC reported in the Table~\ref{Tab:levelsPuO2OH2} shows two close-lying states separated by about \SI{323}{\per\cm}, followed by a state at \SI{2222}{\per\cm}. The analysis of the two lowest states indicates that they correspond to a 50-50 combination of $\mathrm{(1)^{3}A}$ and $\mathrm{(1)^{3}B}$ spin-free states, with a small contribution of the $\mathrm{(2)^{3}B}$ (6\% in each case). The next three states also have strong mixings induced by SOC between triplet states but also singlet excited states of $\mathrm{(1)^{1}A}$ and $\mathrm{(1)^{1}B}$ symmetries, placed at the spin-free level at about \num{5764} and \SI{7444}{\per\cm}, respectively.

\subsubsection{\ce{PuO2} and \ce{PuO2+} Molecules}  

\begin{table*}[htp]
\caption{Comparison of CBS Extrapolated {\dcm}-EOM-CCSD and SO-CASPT2 \ce{PuO2} Vertical Transition Energies ($\Delta E$ in \si{\per\cm}) Computed at the 1.744 and \SI{1.808}{\angstrom} \ce{Pu-O} Distances}
\label{Tab:PuO2_comparison}
\begin{tabular}{l*{3}{S}c*{2}{S}}
\hline
& \multicolumn{3}{c}{d(Pu-O)=\SI{1.744}{\angstrom}}& &\multicolumn{2}{c}{d(Pu-O)=\SI{1.808}{\angstrom}}\\
\cline{2-4}\cline{6-7}
{$\Omega$}& {{\dcm}-EOM-CCSD} & \multicolumn{2}{c}{SO-CASPT2} && {{\dcm}-EOM-CCSD} &{SO-CASPT2}\\
& {$\Delta E$} & {$\Delta E$} & {$\Delta E$\cite{actinide-La-Macchia-PCCP2008-10-7278}} && {$\Delta E$} & {$\Delta E$}\\
\hline
$0_g^+$	&	0	&	0	&	1794	&&	0	&	0	\\
$1_g$	&	2554	&	1438	&	2315	&&	2457	&	2462	\\
$2_g$	&	6366	&	4235	&	4131	&&	5663	&	5904	\\
$0_g$	&	7138	&		&		&&	7134	&		\\
$3_g$	&	9480	&		&		&&	9073	&		\\
$1_g$	&	10098	&		&		&&	9502	&		\\
$0_g$	&	10195	&		&		&&	10415	&		\\
\\											
$1_u$	&	7011	&	1778	&	0	&&	9822	&	5851	\\
$2_u$	&	7225	&	1805	&	535	&&	10057	&	6265	\\
$2_u$	&	7653	&		&		&&	10854	&		\\
$3_u$	&	7955	&		&		&&	11214	&		\\
\\
$4_u$	&	13531	&		&		&&	16236	&		\\
\hline
\end{tabular}
\end{table*}

\begin{figure}[htp]
\centering
\includegraphics[width=\linewidth]{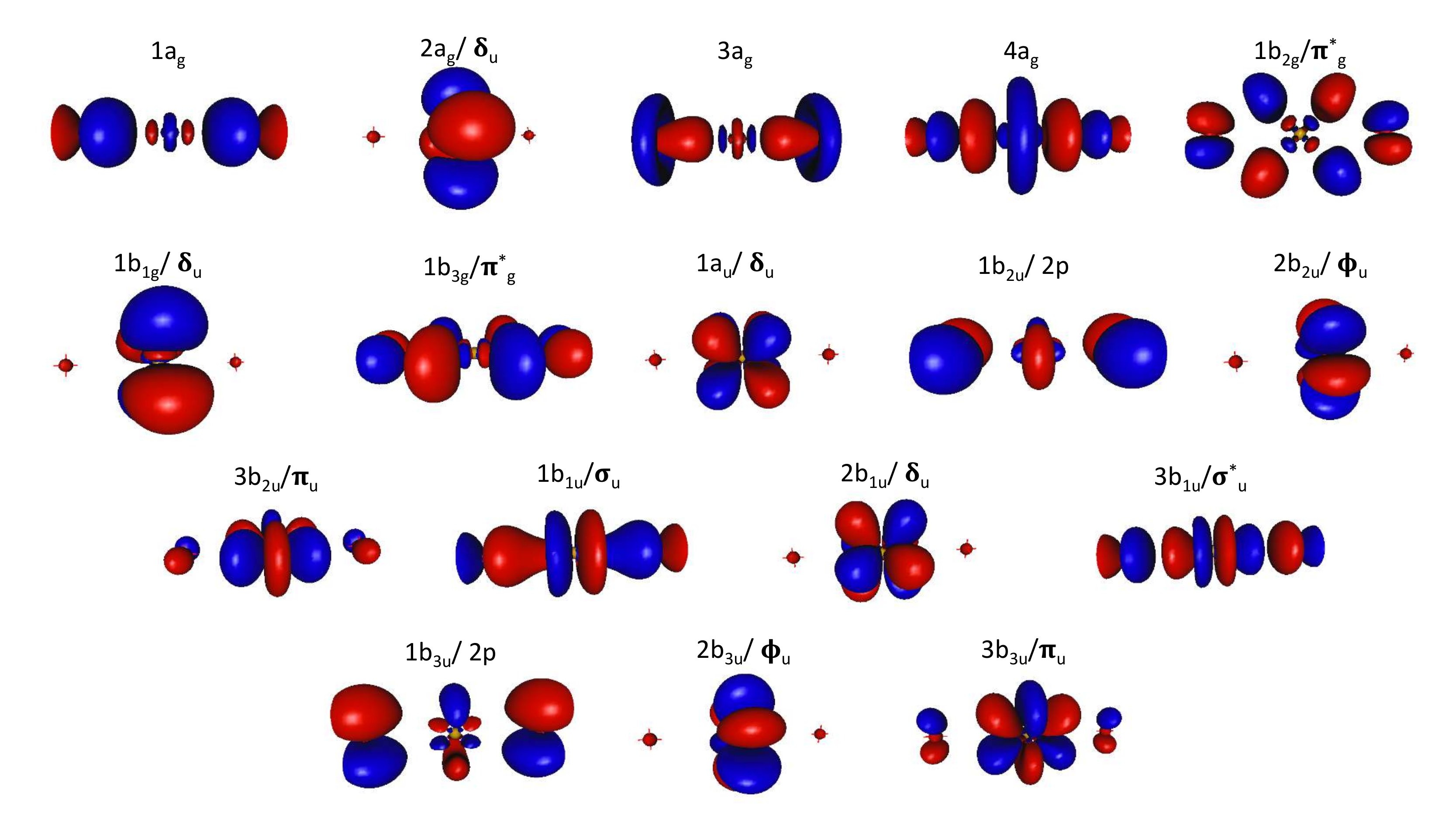}
\caption{Molecular orbitals of \ce{PuO2} at the CASSCF level; \ce{r(Pu-O)}=\SI{1.808}{\angstrom}; Isosurface = \SI{0.05}{\astronomicalunit}.}
\label{Fig:MolOrbPuO2}
\end{figure}

At the {\dcm} level, the \ce{PuO2} ground state is found to be well-described by a closed-shell ($\mathrm{\Omega = 0_g}^+$) determinant, corresponding to occupied $\mathrm{5f_{-3/2; +3/2}}$, $\mathrm{5f_{-5/2; +5/2}}$ spinors. We note that the $T_1$ diagnostic is 0.03, a value very similar to that found for single-reference uranium molecules \cite{actinide-Tecmer-PCCP2011-13-6249}. This allowed us to employ accurate two-component single-reference approaches such as CCSD(T) to determine the optimal \ce{Pu-O} bond length after CBS extrapolation to be \SI{1.808}{\angstrom}, with a corresponding harmonic vibrational symmetric stretching frequency of \SI{791}{\per\cm} (the anti-symmetric stretch and bending frequencies being \num{840} and \SI{170}{\per\cm}, respectively). We note that these results are in excellent agreement with the scalar relativistic CCSD(T) geometry (\SI{1.814}{\angstrom}) and harmonic frequencies (\num{147}, \num{777}, \SI{825}{\per\cm}) reported by Feng and Peterson~\cite{actinide-Feng-JCP2017-147-084108}, who agree with us that \ce{PuO2} ground state has a $\mathrm{0_g^+}$ character with a longer distance than the one (\SI{1.744}{\angstrom}) predicted by La Macchia~{\etal} \cite{actinide-La-Macchia-PCCP2008-10-7278,actinide-Infante-JPC2010-114-6007} at the SO-CASPT2 level, though for chemically different ground state $\mathrm{^5\Phi_u}$.

Furthermore, our {\dcm}-EOM-EE-CCSD calculations (see Table~\ref{Tab:PuO2_comparison}) confirm that, at the CBS-{\dcm}-CCSD(T) equilibrium structure, the $\mathrm{\Omega = 0_g^+}$ state is indeed the ground state: it is sufficiently well-separated  from the lowest-lying states of both g ($\mathrm{\Omega = 1_g}$, by over \SI{2000} {\per\cm}) and u ($\mathrm{\Omega = 1_u}$, by over \SI{5000} {\per\cm}), and  considering a shorter bond length, namely \SI{1.744}{\angstrom}, as was proposed by La Macchia~{\etal} \cite{actinide-La-Macchia-PCCP2008-10-7278} for the ground state from SO-CASPT2 calculations, does not alter this picture. Thus, our results differ qualitatively and quantitatively from ref. \citenum{actinide-La-Macchia-PCCP2008-10-7278}, which found the \ce{PuO2} ground state to be of $\mathrm{\Omega = 1_u}$ symmetry, corresponding to the occupation of the two $\delta_u$, one $\phi_u$ and the 7s orbitals (See Figure~\ref{Fig:MolOrbPuO2}).

From Table~\ref{Tab:PuO2_comparison}, we observe significant variations for the u-states (of nearly \SI{3000}{\per\cm}) when the internuclear distance is shortened from \SI{1.808}{\angstrom} to  \SI{1.744}{\angstrom}, while the g-states remain more or less at the same energies. With this, at this longer distance, the $\mathrm{\Omega = 2_g}$ state becomes lower than the $\mathrm{\Omega = 1_u}$ state.

\begin{table*}[htp]
\caption{Lowest Vertical Transition Energies ($\Delta E$ in \si{\per\cm}) of \ce{PuO2} Computed at the SF-CASPT2 and SO-CASPT2 Levels (d(Pu-O)=\SI{1.808}{\angstrom}) with ANO-RCC-TZVP Basis Sets and Analysis of the Various States}
\label{Tab:levelsPuO2}
\begin{tabular}{ccSl}
\hline
& state  &  {$\Delta E$}   &  character  : \% [orbital(number of electron)]\\
\hline
SF &$\mathrm{(1)^{5}\Sigma_g^+}$ &  0  &  84\% [$\mathrm{1a_u}$/$\mathrm{\delta_u}$(1), $\mathrm{2b_{2u}}$/$\mathrm{\phi_u}$(1), $\mathrm{2b_{1u}}$/$\mathrm{\delta_u}$(1), 2b$_{3u}$/$\mathrm{\phi_u}$(1)] \\
&   $\mathrm{(1)^{5}\Sigma_{g}^-}$   &  4798  &  45\% [$\mathrm{1a_u}$/$\mathrm{\delta_u}$(1),  3b$_{2u}$/$\mathrm{\pi_u}$(1), $\mathrm{2b_{1u}}$/$\mathrm{\delta_u}$(1), 2b$_{3u}$/$\mathrm{\phi_u}$(1)]  \\
& & & 45\% [$\mathrm{1a_u}$/$\mathrm{\delta_u}$(1), $\mathrm{2b_{2u}}$/$\mathrm{\phi_u}$(1), $\mathrm{2b_{1u}}$/$\mathrm{\delta_u}$(1), 3b$_{3u}$/$\mathrm{\pi_u}$(1)]  \\

  &  $\mathrm{(1)^{5}\Phi_{u}}$&  5421  &  85\% [$\mathrm{3a_g}$(1), $\mathrm{1a_u}$/$\mathrm{\delta_u}$(1), $\mathrm{2b_{1u}}$/$\mathrm{\delta_u}$(1), 2b$_{3u}$/$\mathrm{\phi_u}$(1)]\\
  &  $\mathrm{(1)^{5}\Delta_{u}}$ &  7452  &  79\% [$\mathrm{3a_g}$(1), $\mathrm{2b_{2u}}$/$\mathrm{\phi_u}$(1), $\mathrm{2b_{1u}}$/$\mathrm{\delta_u}$(1), 2b$_{3u}$/$\mathrm{\phi_u}$(1)] \\
 &  $\mathrm{(1)^{3}\Sigma_g^-}$ &  10308  & 24\% [$\mathrm{1a_u}$/$\mathrm{\delta_u}$(1), $\mathrm{2b_{2u}}$/$\mathrm{\phi_u}$(2), $\mathrm{2b_{1u}}$/$\mathrm{\delta_u}$(1)] \\
 & & & 24\% [$\mathrm{1a_u}$/$\mathrm{\delta_u}$(1), $\mathrm{2b_{1u}}$/$\mathrm{\delta_u}$(1), 2b$_{3u}$/$\mathrm{\phi_u}$(2)] \\
& & & 19\% [$\mathrm{2b_{2u}}$/$\mathrm{\phi_u}$(1), $\mathrm{2b_{1u}}$/$\mathrm{\delta_u}$(2), 2b$_{3u}$/$\mathrm{\phi_u}$(1)] \\
& & & 16\% [$\mathrm{1a_u}$/$\mathrm{\delta_u}$(1), $\mathrm{2b_{2u}}$/$\mathrm{\phi_u}$(1), 2b$_{3u}$/$\mathrm{\phi_u}$(1)] \\
\hline
SO  & $\mathrm{0_g^+}$ &   0  &  48\%    $\mathrm{(1)^{5}\Sigma_g^+}$,   22\%    $\mathrm{(1)^{3}\Sigma_g^-}$          \\
  & $\mathrm{1_g}$  & 2462  &  67\% $\mathrm{(1)^{5}\Sigma_g^+}$,  15\%    $\mathrm{(1)^{3}\Sigma_g^-}$            \\
 &  $\mathrm{1_u}$& 5851  &  27\%     $\mathrm{(1)^{5}\Phi_{u}}$,  27\%    21\%   $\mathrm{(2)^{5}\Phi_{u}}$,  13\%    $\mathrm{(1)^{5}\Delta_{u}}$,  13\%    $\mathrm{(2)^{5}\Delta_{u}}$ \\
& $\mathrm{2_g}$  & 5904  & 85\%    $\mathrm{(1)^{5}\Sigma_g^+}$  \\
& $\mathrm{2_u}$  & 6265  &  21\%    $\mathrm{(1)^{5}\Phi_{u}}$,  21\%   $\mathrm{(2)^{5}\Phi_{u}}$,  24\%    $\mathrm{(1)^{5}\Delta_{u}}$  \\
&$\mathrm{3_u}$ &	7195	 &	39\%	$\mathrm{(1)^{5}\Phi_{u}}$,	25\%		 $\mathrm{(2)^{5}\Phi_{u}}$		\\
\hline
\end{tabular}
\end{table*}

The SO-CASPT2 transition energies computed at the {\dcm} optimal bond length are reported in Tables~\ref{Tab:PuO2_comparison} and~\ref{Tab:levelsPuO2} and Table~S5 in the Supporting Information. The spin-orbit $\mathrm{0_g^+}$ ground state is composed by the spin-free ground state $\mathrm{(1)^{5}\Sigma_g^+}$ up to 48\% and by the $\mathrm{^3\Sigma_g^-(1)}$ up to 22\%, which lies \SI{10308}{\per\cm} above it. Note that the current vertical transition energies computed at the SO-CASPT2 level are in very good agreement with the {\dcm} energies for the g-states, but the u-manifold come out about \SI{4000}{\per\cm} lower in energy at the SO-CASPT2  level than at the {\dcm}-EOM-CCSD level. The fact that the $\mathrm{0_g^+-1_u}$ energy difference computed at the SO-CASPT2 level of theory (\SI{5851}{\per\cm}) agrees best with the spin-orbit corrected CCSD(T) value by Feng and Peterson (\SI{5421}{\per\cm}) than with the {\dcm}-EOM-CCSD value, suggests that triple corrections have a large contribution to such $\mathrm{g}$ to $\mathrm{u}$ transitions involving a change in Pu orbital occupations. Thus, all levels of theory predict the ground state to have $\mathrm{0_g^+}$ symmetry at both shorter and longer \ce{Pu-O} distances, disagreeing with the proposed $\mathrm{1_u}$ nature of the ground state by La Macchia~{\etal} \cite{actinide-La-Macchia-PCCP2008-10-7278}. However, we note that the electronic state spacing within either the g or u symmetries agrees in both SO-CASPT2 calculations. This finding points out to the importance of the choice of the active space to accurately predict the relative energies of g-states involving mostly nonbonding $\mathrm{5f}$ Pu atomic-centered orbitals, versus the u-states involving the more diffuse Pu $\mathrm{7s}$ orbital. 

Beside calculations including SOC, it is informative to discuss the nature of the \ce{PuO2} ground-state without SOC. Our CASPT2 and UCCSD(T) calculations place the $\mathrm{^5\Sigma_g}^+$ state (($\delta_u)^2(\phi_u)^2$ occupations) lower in energy than the $\mathrm{^5\Phi_u}$ state (($\delta_u)^2(\phi_u)^1(7s)^1$) by \SI{7994}{\per\cm} at a Pu-O distance of \SI{1.808}{\angstrom}, as found in the previous studies \cite{actinide-Archibong-JMST2000-530-165,actinide-La-Macchia-PCCP2008-10-7278}.
With the ANO-RCC AVTZ basis set, the equilibrium \ce{Pu-O} distances are \num{1.818} and \SI{1.808}{\angstrom} at the UCCSD(T) and CASPT2 levels, respectively, distances that are close to the {\dcm}-CCSD(T) result. These values are also in line with the SF-CASPT2 (CAS(12,14)) of La Macchia (d(Pu-O))= \SI{1.792}{\angstrom}), but significantly shorter (by about \SI{0.04/0.05}{\angstrom}) than the estimations of Archibong and Ray \cite{actinide-Archibong-JMST2000-530-165}. The large discrepancies observed with respect to the latter are probably related to the choice of a large-core RECP (78 electrons), which yields bond distances in actinide molecules that are too long \cite{actinide-Batista-JCP2004-121-2144}, while the small-core ECP gives exactly the same distances as the all-electron Douglas-Kroll relativistic approach.

The ground state $\mathrm{(1)^{5}\Sigma_g^+}$ is dominated (84\%) by the configuration in which the two 5$\delta_u$ and two $\phi_u$ orbitals carry one electron each. It corresponds to the description of the ground state of Boguslawski~{\etal} \cite{actinide-Boguslawski-PCCP2017-19-4317} and by Feng and Peterson~\cite{actinide-Feng-JCP2017-147-084108}, and also to the $\mathrm{(1)^{5}\Sigma_g^+}$ state reported by La Macchia~{\etal \cite{actinide-La-Macchia-PCCP2008-10-7278}. The former study places the first excited state at \SI{1800}{\per\cm} corresponding to $\mathrm{^{5}\Phi_{u}}$ state, while in our SF-CASPT2 calculation it appears at a higher energy (\SI{5421}{\per\cm}). We note that in our SF-CASPT2 calculation the spectrum is less dense than in La Macchia's results \cite{actinide-La-Macchia-PCCP2008-10-7278}. Such differences can be explained by the choice of different active spaces, by the fact that we computed a larger number of spin-free states, and most likely by the change in the interatomic \ce{Pu-O} distance. SOC is not expected to change the energy of the $\mathrm{\Omega=0^-_g}$ state, while it lifts the degeneracy of the $\mathrm{^{5}\Phi_{u}}$ state, giving rise to three ($\mathrm{1_u}$, $\mathrm{2_u}$, and $\mathrm{3_u}$) fine structure states, split by \SI{2133}{\per\cm} (See Table~4 of ref. \citenum{actinide-La-Macchia-PCCP2008-10-7278}). Given the small energy gap of \SI{1800}{\per\cm} in La Macchia's \cite{actinide-La-Macchia-PCCP2008-10-7278} study between the $\mathrm{^5\Sigma_g^+}$  and $\mathrm{^5\Phi_u}$ states, the $\mathrm{1_u}$ state becomes the ground state, while in our work, it remains a state of $gerade$ symmetry ($\mathrm{0_g^+}$), unambiguously confirmed by the {\dcm} calculations.

\begin{table*}[htp]
\caption{CBS Extrapolated Vertical and Adiabatic Ionization Potentials (IPs in \si{\electronvolt}) of \ce{PuO2} Computed at the {\dcm}-EOM-IP-CCSD Level. The interatomic distance ($r_e$(Pu-O)= \SI{1.808}{\angstrom}, from {\dcm}-CCSD(T) calculations) was used for the vertical IPs, while the adiabatic IPs correspond to the reported equilibrium $r_e$ distances of \ce{PuO2+} and harmonic symmetric stretching $\omega_e$ frequencies. Experimental and other theoretical values correspond to adiabatic IPs}
\label{Tab:PuO2_IP}
\begin{tabular}{l*4{S}}
\hline
\multicolumn{5}{c}{{\dcm}-EOM-IP-CCSD}\\
State& {$\mathrm{IP_{vert}}$} & {$\mathrm{IP_{adiab}}$} & {$r_e$} & {$\omega_e$}\\
 \hline
$\mathrm{3/2_{u}}$	&	7.67	&	7.07	&	1.697	&	1010  \\
$\mathrm{5/2_{u}}$	&	7.93	&	7.20 &	1.688	&	1035  \\
$\mathrm{1/2_{u}}$	&	9.81	&	9.77	&	1.771	&	  811  \\
$\mathrm{3/2_{u}}$	&    10.80	&    10.79	&	1.831	&	  940  \\
\hline
\multicolumn{5}{c}{Theor. ($\mathrm{IP_{adiab}}$)} \\
{X2C-DC-CCSD(T)} && 5.93 &{ref. \citenum{actinide-Infante-JPC2010-114-6007}} \\
{SO-CCSD(T)} & & 6.93 & {ref. \citenum{actinide-Feng-JCP2017-147-084108}} \\
\hline
\multicolumn{5}{c}{Exp ($\mathrm{IP_{adiab}}$)} \\
&& 9.4\pm0.5 & {ref. \citenum{IP-Rauh-JCP1974-60-1396-1400}}\\
&& 10.1\pm0.1 & {ref. \citenum{actinide-Capone-JPC1999-103-10899}}\\
&& 7.03\pm0.12 & {ref. \citenum{IP-Santos-JPC2002-106-7190}}\\
&& {\textit{6.6}\textsuperscript{\emph{a}}} & {refs. \citenum{actinide-Capone-JPC2005-109-12054,actinide-Gibson-JPC2006-110-4131}}\\
\hline
\end{tabular}

\textsuperscript{\emph{a}} Rejected value according to the comment by Gibson~{\etal} \cite{actinide-Gibson-JPC2006-110-4131} on Capone~{\etal}'s measurements \cite{actinide-Capone-JPC2005-109-12054}.
\end{table*}%

Apart from determining the excitation energies for \ce{PuO2}, the EOM-CCSD method is also useful to investigate the ionization potentials of \ce{PuO2}, for which there are experimental data. Being a relative measure, this can help to validate {\dcm}-EOM-CCSD as a reliable approach with which to benchmark SO-CASPT2. The computed vertical and adiabatic ionization energies (IP; see Table~\ref{Tab:PuO2_IP}) can be grouped into three regions with respect to their energies: two below \SI{8}{\electronvolt}, one between 8  and \SI{10}{\electronvolt} and the last at energies higher than \SI{10}{\electronvolt}. The lowest EOM-IP-CCSD state has $\mathrm{\Omega = 3/2_u}$ symmetry, as found by Feng and Peterson~\cite{actinide-Feng-JCP2017-147-084108}, with an adiabatic IP of \SI{7.07}{\electronvolt}, in very good agreement with the SO-CCSD(T) value (\SI{6.93}{\electronvolt})~\cite{actinide-Feng-JCP2017-147-084108}, and within the error bar of the adiabatic value \SI{7.03\pm0.12}{\electronvolt}, measured by Santos~{\etal} \cite{IP-Santos-JPC2002-106-7190} However, the values reported by Rauh~{\etal} \cite{IP-Rauh-JCP1974-60-1396-1400} and Capone~{\etal} \cite{actinide-Capone-JPC1999-103-10899}, lying at \num{9.4\pm0.5} and \SI{10.1\pm0.1}{\electronvolt}, respectively would match the value we computed for the first $\mathrm{\Omega = 1/2_u}$, leading us to suggest that these experiments may have measured \ce{PuO2+} in different electronically excited states.

The calculated electronic states of \ce{PuO2+} correspond to ionizations from the nonbonding $\mathrm{\delta_u}$ ($\mathrm{\Omega = 3/2_u}$) and $\mathrm{\phi_u}$ ($\mathrm{\Omega = 5/2_u}$), and from the bonding $\mathrm{\sigma_u}$ ($\mathrm{\Omega = 1/2_u}$) and $\mathrm{\pi_u}$ ($\mathrm{\Omega = 3/2_u}$). In the first two cases, as we do not alter the number of bonding spinors, the removal of an electron increases the Pu-O bond energy, as can be seen by the decreased equilibrium bond length and  higher harmonic vibrational stretching frequencies compared to the CCSD values ($r_e$(\ce{Pu-O})=\SI{1.792}{\angstrom}; $\omega_e$=\SI{833}{\per\cm}) for  \ce{PuO2} (by about \SI{0.095}{\angstrom} and 177 \si{\per\cm} for $\mathrm{\Omega = 3/2_u}$, and \SI{0.104}{\angstrom} and \SI{202}{\per\cm} for $\mathrm{\Omega = 5/2_u}$). 

The other two states, on the other hand, can be seen as excited states where the electrons from the bonding  $\mathrm{\sigma_u}$ and $\mathrm{\pi_u}$ spinors are excited to the nonbonding $\mathrm{\delta_u}$ and as such the net bonding interaction is reduced in the molecule, whereas for the second $\mathrm{\Omega = 3/2_u}$ there is an increase in bond length and in frequency in comparison to the \ce{PuO2} CCSD(T) values (by \SI{0.039}{\angstrom} and \SI{107}{\per\cm}). Finally, for the $\mathrm{\Omega = 1/2_u}$, we observe a slight decrease in bond length (of \SI{0.021}{\angstrom}) together with a slight decrease in vibrational frequency (of \SI{21}{\per\cm}) in comparison to the \ce{PuO2} CCSD(T) values.

Taken together our {\dcm}-EOM-CCSD calculations show it can be used to assess the SO-CASPT2 method for energy differences.

\subsection{Thermodynamic Data}\label{Sec:Thermo}
All of the results so far have demonstrated the accuracy and reliability of the {\dcm}-CCSD approach and the relative accuracy of SO-CASPT2. Now, we can proceed with the determination of the thermodynamic data. Recently, a thorough assessment of the thermodynamic properties of gaseous plutonium oxides has been done by Konings~{\etal} \cite{actinide-Konings-JPCRD2014-43-013101}. The derived functions (entropy and heat capacity) have been calculated on the basis of the most reliable data available in the literature. Nevertheless, the authors pointed out a number of missing data, especially regarding the contributions of electronically excited states. 

In the following, from results discussed before and the related molecular parameters and electronic states (see Tables~S7--S9 in the Supporting Information), the heat capacity and the standard entropy of \ce{PuO2(g)},  \ce{PuO3(g)} and \ce{PuO2(OH)2(g)} are thus derived and compared to those selected by Konings~{\etal} \cite{actinide-Konings-JPCRD2014-43-013101} or calculated by Ebbinghaus \cite{actinide-Ebbinghaus-TR-2002}.  Regarding the standard heat of formation, an analysis of transpiration measurements made by Krikorian~{\etal} \cite{actinide-Krikorian-JNM1997-247-161} was performed to extract values of \ce{PuO3(g)} and \ce{PuO2(OH)2(g)}, whereas for \ce{PuO2(g)}, our results are analyzed in the light of Konings' assessment \cite{actinide-Konings-JPCRD2014-43-013101}.

From the discussions on the previous sections of the electronic structures of the different Pu species, one will not be surprised to see that the single-reference  methods, DFT and CC, fail at describing standard heats of formation of \ce{PuO2(OH)2} and \ce{PuO3}, while it should be in principle applicable for \ce{PuO2}. For the latter, data obtained by coupled-cluster calculations are therefore used to establish its thermodynamic properties whereas regarding the other species, multireference results are used.  

\subsubsection{\ce{PuO2(g)}}
Let us first discuss the thermodynamic functions of \ce{PuO2}. Regarding the standard entropy at room temperature, an analysis of the sources of discrepancies between our calculation and up-to-date data from the literature reveals that vibrational and electronic partition functions are in disagreement (see Table~S10 in the Supporting Information).

In the thermodynamic review of Konings~{\etal} \cite{actinide-Konings-JPCRD2014-43-013101}, although information on the \ce{PuO2} electronic states could have been taken from La Macchia~{\etal} \cite{actinide-La-Macchia-PCCP2008-10-7278}, the contributions to the entropy of the \ce{PuO2} electronic states were estimated by considering the crystal-field split $\mathrm{5f^2}$ energy levels of \ce{PuF4(cr)}, which are densely spaced, thus leading to a significant overestimation of the electronic contributions to entropy. Regarding the vibrational contributions, Konings~{\etal} \cite{actinide-Konings-JPCRD2014-43-013101} used the harmonic frequencies computed at the scalar relativistic CCSD(T) level by Archibong and Ray \cite{actinide-Archibong-JMST2000-530-165}. Our {\dcm}-CCSD(T) harmonic stretching frequencies (791 and \SI{840}{\per\cm}) are very close to their CCSD(T) values (792 and \SI{828}{\per\cm}) and to the DKH3-CCSD(T) values of Feng and Peterson~\cite{actinide-Feng-JCP2017-147-084108} (\num{777} and \SI{825}{\per\cm}), the bending frequency comes out slightly higher in our calculation (\SI{170}{\per\cm}) than in previous reports (\SI{106}{\per\cm}\cite{actinide-Archibong-JMST2000-530-165}, and \SI{147}{\per\cm}\cite{actinide-Feng-JCP2017-147-084108}), probably due to the inclusion of spin-orbit coupling. As a result our thermodynamics function differs from that proposed by Konings~{\etal} \cite{actinide-Konings-JPCRD2014-43-013101} Unfortunately, except for the asymmetric stretching mode  by infrared absorption spectroscopy \cite{actinide-Green-JCP1978-69-544} of \ce{PuO2} in Ar and Kr matrices (794.25 and \SI{786.80}{\per\cm}), the other vibrational frequencies are unknown, making it impossible to assess the accuracy of the currently reported quantum chemical frequencies. Our calculated standard entropy at room temperature is thus equal to
\begin{align}
\state[pre=,subscript-right={\SI{298}{\kelvin}}]{S}=\SI{262.9}{\joule\per\kelvin\per\mol},
\end{align} 
which is lower by about \SI{16}{\joule\per\kelvin\per\mol} than the value in the assessment of  Konings~{\etal} (\SI{278.7}{\joule\per\kelvin\per\mol}) \cite{actinide-Konings-JPCRD2014-43-013101}. 
An analogous discussion can be made as to the effect of the \ce{PuO2} electronic states on the heat capacity (see Figure~S1 in the Supporting Information). At room temperature, Konings~{\etal} \cite{actinide-Konings-JPCRD2014-43-013101} reported a $\state[pre=,subscript-right=p]{C}(T)$ value higher than ours by tens of \si{\joule\per\kelvin\per\mol}, due to their overestimated excited-state contributions. The heat capacity function of the present work (in \si{\joule\per\kelvin\per\mol}) is expressed as
\begin{align}
\state[pre=,subscript-right=p]{C}(T)  &= 42.7970+\num{3.5713e-2} (T/K)\nonumber\\
&\num{-1.2077e-5} (T/K)^2-\num{2.0271e5} (T/K)^{-2}
\end{align} 
in the 298.15--\SI{1200}{\kelvin} temperature range, and becomes: 
\begin{align}
\state[pre=,subscript-right=p]{C}(T) &= \num{54.1604} + \num{1.2436e-2} (T/K)\nonumber\\
&\num{-1.7456e-6} (T/K)^2 + \num{2.2335e6} (T/K)^{-2}
\end{align}
in the \num{1200}--\SI{3000}{\kelvin} range.

Coming now to the standard enthalpy of formation of \ce{PuO2}, we first note that the value we report in Table~\ref{Tab:PuOx_Hform} from B3LYP calculations including the spin-orbit $+\Delta E_{SO}$ correction is about 25\% off (\SI{100}{\kJ\per\mol}) the experimental value. A change in the density functional will not ensure a better agreement with experiment as observed previously for actinide systems \cite{actinide-Kovacs-CR2015-115-1725} or for transition metal molecules \cite{transiton-metal-Miradji-JPC2015-119-4961-4971}. Thus, we will rely on wave-function-based approaches to propose revised values of {\Hform}({\Tref}), and will take the CBS corrected {\dcm}-CCSD(T) result to be the best estimate for the standard heat of formation of \ce{PuO2(g)}, yielding \begin{align}
 \enthalpy*(f){}({\Tref}) = \SI{-449.5\pm8.8}{\kJ\per\mol}.
\end{align}
In addition, by combining the {\dcm}-CCSD(T) standard enthalpy of formation of \ce{PuO2} with the  {\dcm}-EOM-IP-CCSD value of its adiabatic ionization potential IE(\ce{PuO2}), we can predict the standard enthalpy of formation of \ce{PuO2+} through the relation 
\begin{align}
\enthalpy*(f){}[\ce{PuO2+}] =&  \enthalpy*(f){}[\ce{PuO2}] + IE[\ce{PuO2}]+\state[pre=]{H}(\Tref)[\ce{PuO2+}] -  \state[pre=]{H}(\Tref)[\ce{PuO2}]\nonumber\\
\enthalpy*(f){}[\ce{PuO2+}] =&  \SI{234.9\pm8.8}{\kJ\per\mol}.
\end{align}

The SO-UCCSD(T) value (\SI{-411.2\pm6.6}{\kJ\per\mol}) for \enthalpy*(f){}[\ce{PuO2}] roughly agrees with both the latter and experiment, but it is lower than the CBS-{\dcm}-CCSD(T) value (\SI{-449.5\pm8.8}{\kJ\per\mol}) by about \SI{38}{\kJ\per\mol}. The importance of the perturbative triple contributions in the electronic correlation treatment is highlighted by comparing these results to the CCSD results (see Table~\ref{Tab:PuOx_Hform}), as the one-step and two-step models show differences of \num{40} and \SI{58}{\kJ\per\mol}, respectively.
The difference between the SO-UCCSD(T) and {\dcm}-CCSD(T) methods could be explained by  a possible underestimation of the SOC contribution in the two-step approach, yielding a {\Hform}({\Tref}) value higher by about \SI{30}{\kJ\per\mol}.
The SO-CASPT2 value ({\Hform}({\Tref})=\SI{-413.7\pm 18.3}{\kJ\per\mol}) reported in Table~\ref{Tab:PuOx_Hform} deviates by about \SI{36}{\kJ\per\mol} from both {\dcm}-CCSD(T) and experiments and by about \SI{2}{\kJ\per\mol} from the SO-UCCSD(T) value, therefore showing a fair agreement with the most accurate methods. This is mirrored in the closeness of the \ce{PuO2} atomization energies (reverse of reaction R2), reported in Table~S11 in the Supporting Information. With this we concluded that perturbative approaches ({\dcm}-CCSD(T) and SO-CASPT2) yield reasonable results, and thus could be used for other systems, though with larger uncertainties for SO-CASPT2 than for {\dcm}-CCSD(T).

Going back to the available literature data, the {\Hform}({\Tref}) computed with the {\dcm}-CCSD(T) approach is in almost perfect agreement with the available results that correspond to the highest values for the formation enthalpy in Table~\ref{Tab:PuOx_Hform} \cite{actinide-Konings-JPCRD2014-43-013101,actinide-Gotcu-Freis-JCT2011-43-1164}. The SO-CASPT2(4,4) value exhibits a rather good agreement with the selected one in the Konings assessment \cite{actinide-Konings-JPCRD2014-43-013101} (the value has been obtained by a second law method applied to experiments published in ref. \cite{actinide-Gotcu-Freis-JCT2011-43-1164}). This is motivated by the statement that  the values obtained by third or second law analysis are not fully in agreement \cite{actinide-Gotcu-Freis-JCT2011-43-1164,actinide-Konings-JPCRD2014-43-013101} revealing a lack of consistency of experimental data. 

Therefore, in future studies, experimental data can be scrutinized in regard to the present thermodynamic functions of \ce{PuO2}. The crosscheck between data from molecular parameters and standard heat of formation from quantum chemistry calculations as well as values from experiments calls for further assessments.

\subsubsection{\ce{PuO3(g)}}

As highlighted in the case of the \ce{PuO2(g)} molecule, the uncertainties related to the entropy and heat capacity functions stem from the electronic partition functions. Ebbinghaus \cite{actinide-Ebbinghaus-TR-2002} and Konings~{\etal} \cite{actinide-Konings-JPCRD2014-43-013101}, both used the electronic spectra of \ce{PuO2^{2+}} to estimate that of \ce{PuO3} \cite{actinide-Infante-JCP2006-125-074301,actinide-Eisenstein-JRNBS1966-70A-165}. The latter approximation is found here to be strong, even if both our calculations and the studies by Konings~{\etal} \cite{actinide-Konings-JPCRD2014-43-013101} and Ebbinghaus \cite{actinide-Ebbinghaus-TR-2002} consider that the ground state is dominated by triplet spin-free states. Indeed, while Konings~{\etal} only considered four states from the calculations of Infante~{\etal}'s calculations \cite{actinide-Infante-JCP2006-125-074301}, we find seven states below \SI{8000}{\per\cm}. 

Regarding the vibrational contributions, the estimate used by these authors and based on an analogy with \ce{UO3(g)}, seems to be acceptable because of a fortuitous good agreement, as highlighted by our calculations. The improvements of the electronic partition function lead to a weak decrease (see Table~S10 in the Supporting Information) of the standard entropy at room temperature, and is thus equal to 
\begin{align}
\state[pre=,subscript-right={\SI{298}{\kelvin}}]{S} = \SI{310.98}{\joule\per\kelvin\per\mol}, 
\end{align}
and the heat capacity function of \ce{PuO3(g)} (in \si{\joule\per\kelvin\per\mol}) is
\begin{align}
\state[pre=,subscript-right=p]{C}(T) &= 43.0757+\num{1.0927e-1} (T/K) \nonumber\\
&\num{-7.1581e-5} (T/K)^2 - \num{7.1469e4} (T/K)^{-2}
\end{align} 
for the 298.15--\SI{550}{\kelvin} temperature range, and 
\begin{align}
\state[pre=,subscript-right=p]{C}(T) &= 98.1101-\num{1.1768e-2} (T/K)\nonumber\\
&+\num{4.5533e-6} (T/K)^2-\num{3.5486e6} (T/K)^{-2}.
\end{align}
in the \num{550}--\SI{2400}{\kelvin} range.

The only available experimental data of \ce{PuO3(g)} for which its standard heat of formation has been obtained comes from the transpiration method of plutonium oxide under an oxygen atmosphere \cite{actinide-Krikorian-JNM1997-247-161}. The involved species is postulated from an analogy with uranium, and under the experimental conditions (under a purely oxygen flow), \ce{PuO3} is expected to be the major volatile species (under a purely oxygen flow). In this type of apparatus, the partial pressure of target species is obtained by an indirect method (measure of transported masses), which may cause a bias in the present case due to a potential contamination of the results by the ash transport (pointed out by the authors of ref. \citenum{actinide-Krikorian-TechReport114774-1993}). Nevertheless, one set of measures has been obtained with another apparatus configuration in which a silica wool glass filter has been added between the crucible and the collector tube (in a region where the temperature remains high).

Strangely, these obtained data are excluded because the following postulate has been made by authors:  the lowest plutonium volatility is due to an interaction between its vapor and silica wool. Therefore, the standard heat of formation retained by the authors (\SI{-562.8}{\kJ\per\mol}) \cite{actinide-Krikorian-JNM1997-247-161} has been obtained with the set without a silica wool filter. However, one can make the assumption that the filter traps the ash. As the filter is located in a hot zone, in the vicinity of the crucible, one can postulate that only vapor species pass through the filter without modifying the equilibrium. Thus, the data set obtained with a filter may be more accurate than that without a silica wool filter.    

Taking advantage of our improved entropy and heat capacity for the \ce{PuO3} molecule, we can reanalyze the experimental data using our \textit{ab initio} data. The treatments by the third law method with the proper Gibbs energy functions of \ce{PuO2(c)} \cite{actinide-Konings-JPCRD2014-43-013101} and one from our adopted data for \ce{PuO3(g)} lead to the following \enthalpy*(f){}{} values: \SI{-572\pm 12.1}{\kJ\per\mol} from the data set without a silica wool filter and \SI{-534.0\pm 18.5}{\kJ\per\mol} with it\footnote[2]{the uncertainties are given within a level of confidence of 95 percent}. The former slightly differs from the original value (\SI{-562.8}{\kJ\per\mol}) \cite{actinide-Krikorian-JNM1997-247-161}, due to our improvement of the standard entropy for \ce{PuO3(g)}. We propose that the standard heat of formation computed within the SO-CASPT2 approach can be considered as the best estimated data  
\begin{align}
\enthalpy*(f){}{}(\Tref)=\SI{-553.2\pm27.5}{\kJ\per\mol}.
\end{align}
The predicted theoretical  \enthalpy*(f){}{}(\Tref) is in the same range as the values reported in Table~\ref{Tab:PuOx_Hform} and is in agreement with the new assessments of experimental data proposed here.

\subsubsection{\ce{PuO2(OH)2(g)}}
Ebbinghaus' work  reports the evaluation of the entropy and heat capacity of the plutonium oxy-hydroxide \cite{actinide-Ebbinghaus-TR-2002}. To compute the electronic contribution to the thermodynamics function of \ce{PuO2(OH)2(g)}, Ebbinghaus also used, as for \ce{PuO3(g)}, the electronic spectrum of \ce{PuO2^{2+}}, but the main discrepancy in the standard entropy is due to the treatment of the vibrational frequencies. Actually, in Ebbinghaus' work a full analogy to \ce{UO2(OH)2(g)} was assumed, and the simulation treated the two torsional modes (for this molecule, internal rotations of OH groups) with the hindered rotor approach instead of harmonic oscillators. Nonetheless, in a recent study related to the thermodynamic properties of \ce{UO2(OH)2(g)} \cite{actinide-Konings-JNM2017-496-163},  no specific treatment has been made  because it has been pointed out as an unreliable approach for these molecules in a previous review \cite{actinide-Guillaumont-Book2003}. Therefore, similar behavior could be expected for \ce{PuO2(OH)2(g)} molecule: for reasons of computational cost, internal rotations are treated as other vibrations. Finally, for all these reasons, we prefer to rely on our calculations and we propose the following standard entropy of \ce{PuO2(OH)2(g)} at room temperature:
\begin{align}
\state[pre=,subscript-right={\SI{298}{\kelvin}}]{S} = \SI{355.74}{\joule\per\kelvin\per\mol},
\end{align}
which is lower by \SI{22.1}{\joule\per\kelvin\per\mol} than Ebbinghaus' data (\SI{377.8}{\joule\per\kelvin\per\mol}). The heat capacity function (in \si{\joule\per\kelvin\per\mol}) is: 
\begin{align}
\state[pre=,subscript-right=p]{C}(T) &=119.4872+\num{3.3686e-2} (T/K) \nonumber\\
&-\num{9.5789e-6} (T/K)^2-\num{1.6030e6} (T/K)^{-2}
\end{align} 
in the \num{298.15}--\SI{1000}{\kelvin} range, and 
\begin{align}
\state[pre=,subscript-right=p]{C}(T) &=134.27+\num{1.5303e-2} (T/K)\nonumber\\
&-\num{2.0559e-6} (T/K)^2-\num{5.5254e6} (T/K)^{-2}
\end{align}
in the \num{1000}--\SI{3000}{\kelvin} higher temperature range.

With respect to the standard heat of formation, the same drawbacks as for the \ce{PuO3} molecule are pointed out regarding the experimental measurements carried out by Krikorian~{\etal} \cite{actinide-Krikorian-JNM1997-247-161}. Additional issues may increase the uncertainty in this case, as the authors had to do some estimates related to water weighting in order to measure the steam pressure in the apparatus \cite{actinide-Krikorian-JNM1997-247-161}. Thus, our calculations, which are not completely off from the reported experimental estimation (\SI{-1018.2\pm3.3}{\kJ\per\mol}) \cite{actinide-Krikorian-JNM1997-247-161},  can be considered as the most reliable ones. Our best estimate (SO-CASPT2) for the standard heat of formation thus is
\begin{align}
 \enthalpy*(f){}{}(\Tref)=\SI{-1012.6\pm38.1}{\kJ\per\mol}.
\end{align}

\begin{table*}[htp]
\caption{Standard Enthalpies of Formation ({\Hform}({\Tref}) in \si{\kJ\per\mol}) of \ce{PuO2}, \ce{PuO2(OH)2}, \ce{PuO3} in the Gas Phase Calculated at Various Level of Theory and Extrapolated to the Complete Basis Set (CBS) limit and Obtained from the Average over the Six Teactions Listed in Table~\ref{Tab:reaction}, except for {\dcm}-CCSD(T) for Which the Average only Includes Four Reactions R3--R6. $\Delta E_{SO}$ represents the spin-orbit contribution to the enthalpy, which is included in all SO-method results. The uncertainties $\mathrm{\Delta}${\Hform} correspond to 95\% confidence intervals are computed with the formulas~\cite{thermo-Ruscic-JPCRD2005-34-573} given in the Supporting Information}
\label{Tab:PuOx_Hform}
\begin{tabular}{
	l
	S[detect-weight,mode=text]
	S[detect-weight,mode=text]
	S[detect-weight,mode=text]}
\hline
Method & {\ce{PuO2}} &{\ce{PuO3}}& {\ce{PuO2(OH)2}} \\
\hline 

$\Delta E_{SO}$\textsuperscript{\emph{a}}& 34.4 & 59.0 &46.9  \\
SO-B3LYP\textsuperscript{\emph{b}} &  -318.4 \pm 20.6 & -405.7 \pm 22.5& -801.9\pm30.3 \\
SO-UCCSD & -353.3 \pm 16.2 & {NC} &  {NC}\\
SO-UCCSD(T)& -411.2\pm 6.6 & {NC} & {NC}\\
{\dcm}-CCSD\textsuperscript{\emph{c}} &-409.3\pm14.9& {NC} & {NC} \\
{\dcm}-CCSD(T)\textsuperscript{\emph{c}} &\bfseries -449.5\pm8.8 & {NC} & {NC} \\
SO-CASPT2& -413.7\pm 18.5&\bfseries -553.2	\pm27.5& \bfseries -1012.6\pm38.1 \\
\hline
Experiment &\num{-410\pm20} \cite{actinide-Cordfunke-JPE1993-14-457}; \num{-428.7 \pm 7} \cite{actinide-Gotcu-Freis-JCT2011-43-1164}&  -562.8 \pm 5 {\cite{actinide-Krikorian-JNM1997-247-161}}&-1018.2 \pm 3.3 {\cite{actinide-Krikorian-JNM1997-247-161}}\\
& \num{-440\pm 7} \cite{actinide-Gotcu-Freis-JCT2011-43-1164}; \num{-412 \pm 20} \cite{actinide-Glushko-Book1978} & -567.6 \pm 15 {\cite{actinide-Konings-JPCRD2014-43-013101}}&  \\
&-428.7 \pm 20 {\cite{actinide-Konings-JPCRD2014-43-013101}}  \\
\hline
\end{tabular}

\textsuperscript{\emph{a}} $\Delta E_{SO}$ is the estimate of the SO contribution;
\textsuperscript{\emph{b}} The energies of reactions were estimated with a RECP and the associated basis set of triple $\zeta$ quality for the plutonium. Thus, no CBS was performed;
\textsuperscript{\emph{c}} Uncertainties over four reactions $\mathrm{R_3}$--$\mathrm{R_6}$;
\end{table*}

\subsection{Thermodynamic Equilibrium Calculations}

From an experimental viewpoint, the existence of \ce{PuO3(g)} remains an open issue. The most recent analysis, involving mass spectrometric studies, performed on the plutonium vapor phase in the Pu+O system, has not detected the \ce{PuO3+} vaporous species in the apparatus, whatever the vacuum or oxidative environment \cite{actinide-Gotcu-Freis-JCT2011-43-1164}. However, the authors concluded that dedicated experiments are needed to state the existence of this vapor species especially due to the previous study by Ronchi where \ce{PuO3} was briefly detected \cite{Ronchi-JNM2000}.

To investigate the chemical equilibrium of the vapor phase over the plutonium oxygen phase with the current thermodynamic data, thermodynamic equilibrium computations were performed within the Nuclea Toolbox software \cite{nucleatoolbox} and the self-consistent MEPHISTA (version 17$\_$1) database designed for nuclear fuel \cite{mephista}. Isochoric calculations in which the oxygen amount is varying (associated with constant plutonium quantity) are made to show the plutonium vapor trend versus the oxygen potential. The simulation was run at \SI{2270}{\kelvin} to reproduce the experiment under oxidative conditions, carried out by Gotcu-Freis~{\etal} \cite{actinide-Gotcu-Freis-JCT2011-43-1164}. Attention should be focused at the oxygen potential equal to about \SI{-130}{\kJ\per\mol}, corresponding to the oxygen pressure of \SI{100}{\pascal} applied during the experiment. In order to perceive the effect of the newly determined thermodynamics functions, a first calculation with the original thermodynamic database (MEPHISTA-17$\_$1) was performed (Figure~\ref{Fig:Pp17}). At an oxygen potential of \SI{-130}{\kJ\per\mol} the gaseous phase is mainly composed of \ce{PuO2} though its partial pressure remains low. In addition, the contribution to the gaseous phase of \ce{PuO} increases as the oxygen's chemical potential decreases.

\begin{figure}[htp]
\begin{subfigure}[t]{0.5\textwidth}
\centering
\includegraphics[width=\linewidth]{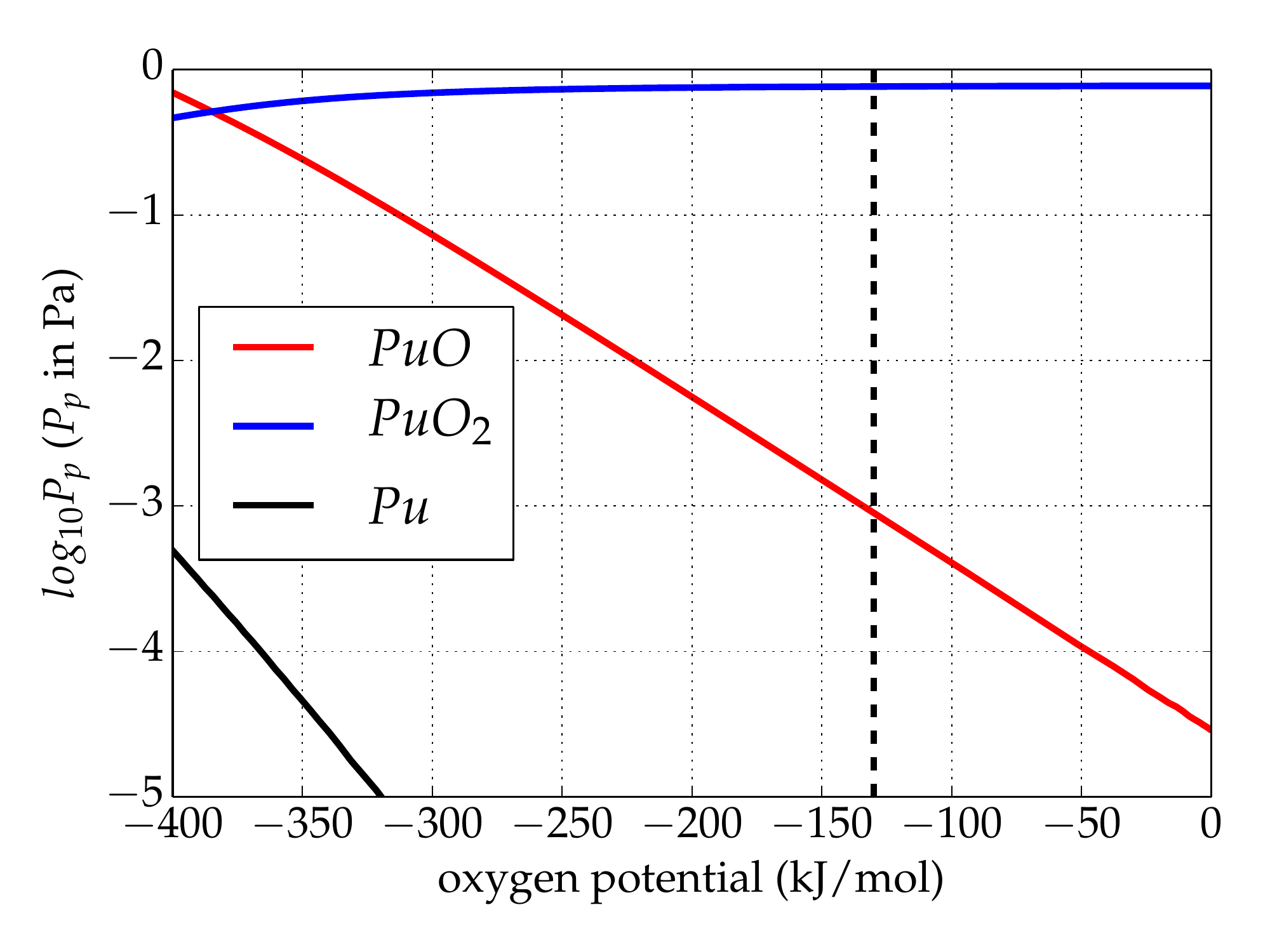}
\caption{MEPHISTA-17$\_$1}\label{Fig:Pp17}
\end{subfigure}
\begin{subfigure}[t]{0.5\textwidth}
\centering
\includegraphics[width=\linewidth]{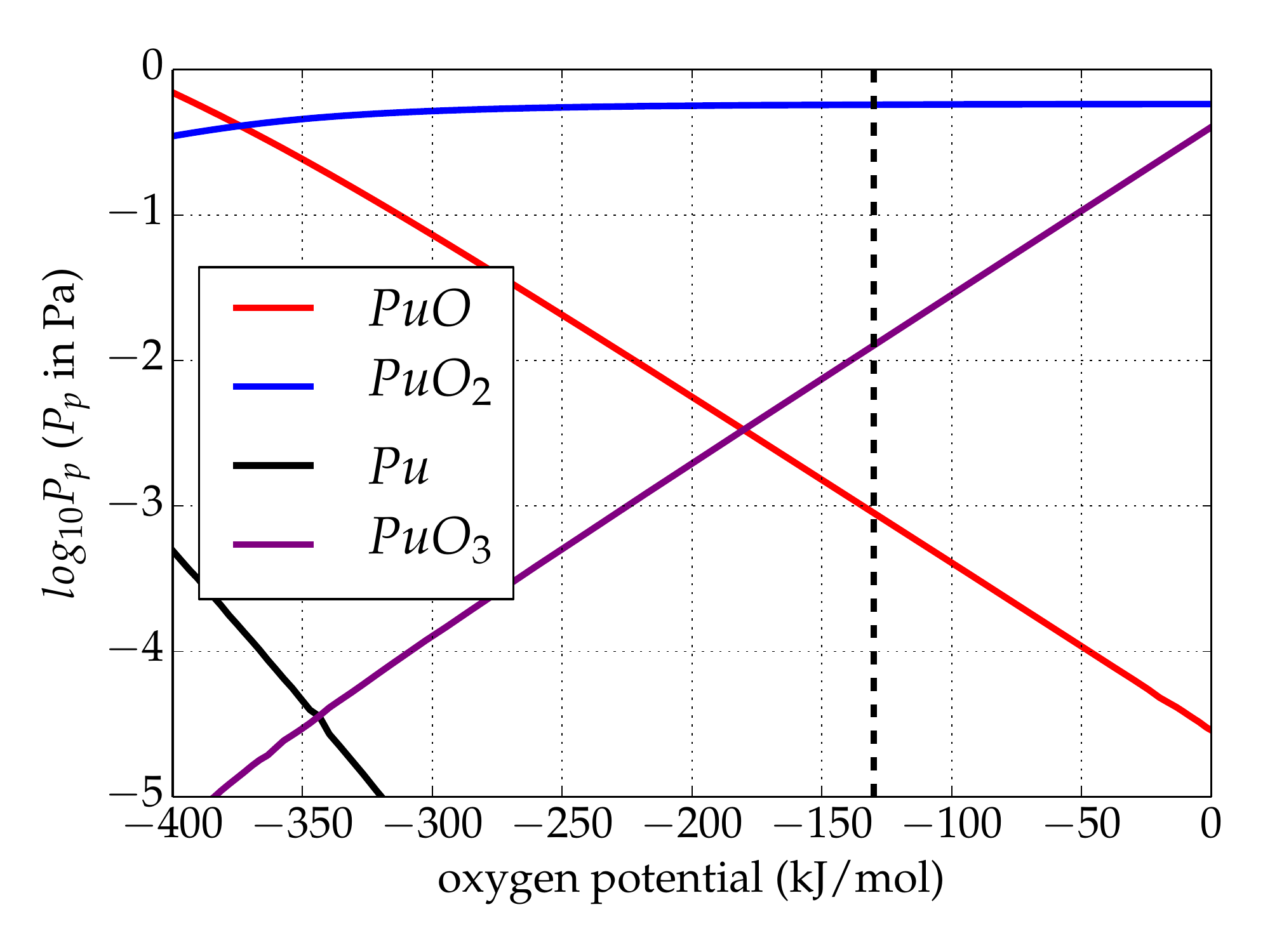}
\caption{Modified database}\label{Fig:Ppnew}
\end{subfigure}
\caption{Thermodynamic equilibrium calculations : focus on the partial pressures of plutonium species at \SI{2270}{\kelvin} versus the chemical potential of oxygen. The dashed line corresponds to an oxygen pressure of \SI{100}{\pascal} (\SI{-130}{\kJ\per\mol}).}
\label{Fig:Pp}
\end{figure}

Conversely, with the updated database (Figure \ref{Fig:Ppnew}), i.e. \ce{PuO3(g)} is added and the Gibbs function of \ce{PuO2(g)} is modified according to the present work. Unlike the first simulation, \ce{PuO3} appears in the gaseous phase up to low oxygen potentials, whereas its partial pressure is higher than that of \ce{PuO} for an oxygen potential of about  \SI{-130}{\kJ\per\mol}. This appears to contradict the analysis of Gotcu-Freis~{\etal} \cite{actinide-Gotcu-Freis-JCT2011-43-1164}, in which \ce{PuO} is detected whereas the same is not true for \ce{PuO3}.
The dissociation energy of \ce{PuO3} to \ce{PuO2+} is \SI{1032}{\kJ\per\mol}, calculated from the reviewed data of the ionization energy of \ce{PuO2} (\SI{7.07}{\electronvolt}). The SO-CASPT2 calculations allow us to propose the value of \SI{9.2}{\electronvolt} (\SI{887.5}{\kJ\per\mol}) for the adiabatic ionization energy for \ce{PuO3}, about \SI{0.9}{\electronvolt} lower that that predicted (\SI{10.1}{\electronvolt}) by scalar relativistic DFT calculations~\cite{actinide-Gao-AC2004-62-454}. The \ce{PuO3} ionization energy is just below the dissociation energy and therefore suggests that the fragmentation of \ce{PuO3} to \ce{PuO2+} could be an important process and explains why \ce{PuO3} is not detected.
Finally, even if the temperature is high, these simulations show that the total partial pressure of plutonium remains rather low, whatever the database. At high oxygen potentials, a temperature drop should promote the plutonium trioxide over the dioxide with nevertheless a partial pressure lower than that obtained at \SI{2270}{\kelvin}. 

To extend the investigation to an oxygen/steam environment case, thermodynamic equilibrium calculations were also done under isobaric condition (standard pressure) with a constant amount of oxygen while a variable hydrogen inventory was used in order to shift the oxygen potential. The partial pressures of gaseous plutonium species versus the oxygen potential were calculated at the lowest possible temperature (\SI{1500}{\kelvin}) at which gaseous plutonium is present, and in accordance with the simulation temperature used in the experiments dedicated to Pu(VI) volatility performed by Krikorian~{\etal} \cite{actinide-Krikorian-JNM1997-247-161} and Hubener~{\etal} \cite{actinide-Hubener-RA2008-96-781}. At this temperature, the plutonium volatility is very low: i.e., < 10$^{-11}$ bar whatever the oxygen potential. Under a purely oxygen atmosphere, as expected, only \ce{PuO3} appears, but if steam is present, the \ce{PuO2(OH)2} species exists and rapidly becomes the main gaseous species when the steam fraction rises (see Fig.~\ref{Fig:H-O}).  
 
\begin{figure}[htp]
\centering
\includegraphics[width=\linewidth]{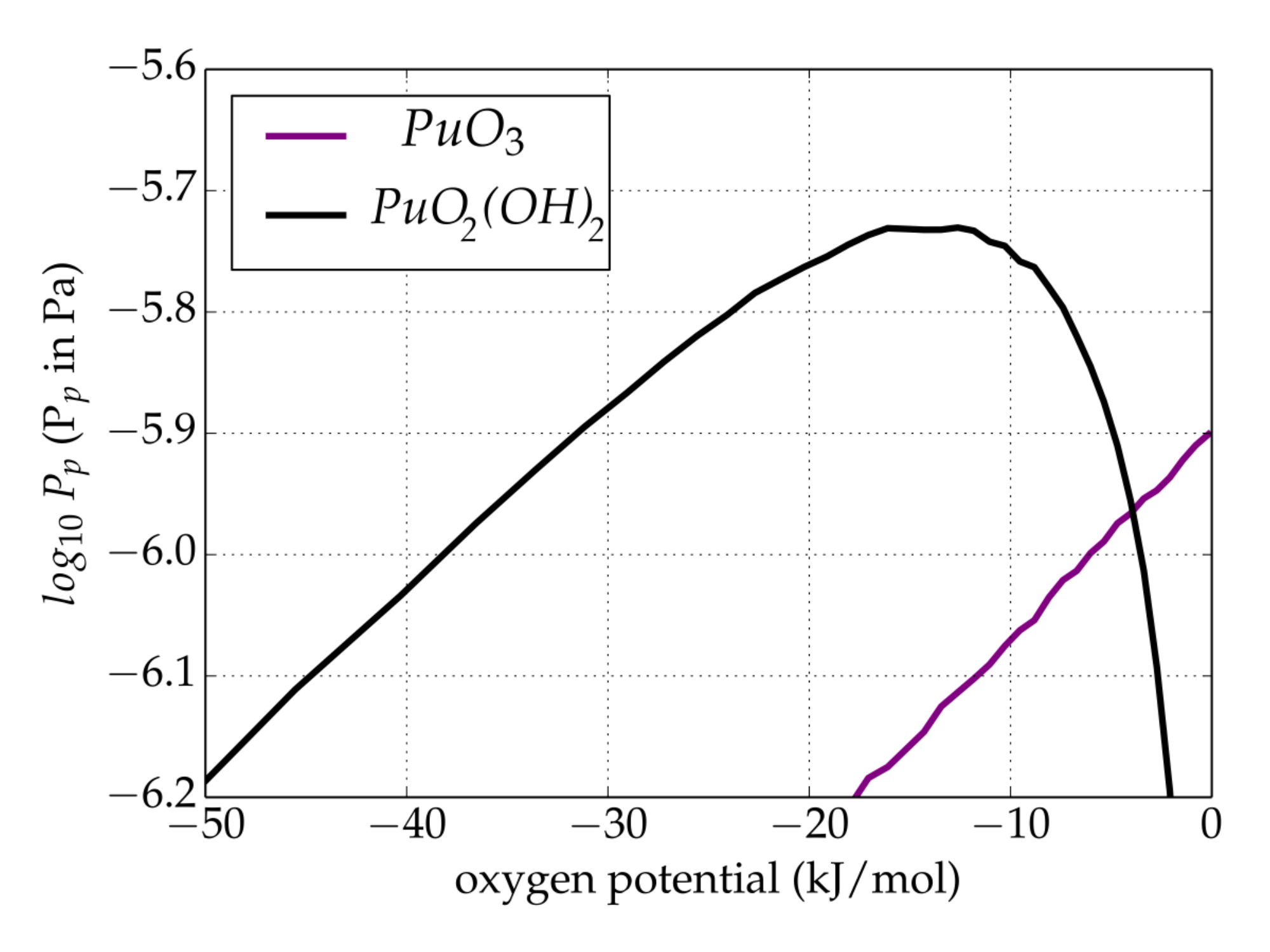}
\caption{Focus on the partial pressures of plutonium species at \SI{1500}{\kelvin} versus the chemical potential of oxygen under standard pressure conditions for the Pu-O-H system.}
\label{Fig:H-O}
\end{figure}

\section{Conclusions}\label{Sec:conclusions}
In conclusion, this work proposes a complete revision of the thermodynamics functions of the main gaseous species of the \ce{Pu-O-H} system, namely \ce{PuO2}, \ce{PuO3}, and \ce{PuO2(OH)2}, using accurate relativistic correlated quantum chemical calculations extrapolated to the complete basis set limit. This updated database is helpful to predict the most important thermophysical properties for nuclear safety and risk analysis: namely, the vapor pressure. Thermodynamic equilibrium calculations with the updated database performed at high temperature disclose new aspects of the volatility of plutonium under accidental conditions. They not only confirm the predominance of \ce{PuO2} under an oxygen atmosphere, but also validate the conjecture of Ronchi~{\etal} \cite{Ronchi-JNM2000} and Krikorian~{\etal} \cite{actinide-Krikorian-JNM1997-247-161} regarding the formation of the molecule \ce{PuO3(g)} at high oxygen potentials. However, all reported quantum chemical calculations suggest that the stability of \ce{PuO3(g)} might be hampered by fragmentation processes to lower-valent oxides. In the presence of steam, plutonium volatilizes into two competing Pu(VI) gaseous forms, \ce{PuO3(g)}, and \ce{PuO2(OH)2(g)}, though their partial pressures are low: $i.e.$, < 10$^{-11}$ bar. This study increases the knowledge of the effective volatility of plutonium and calls for caution assessments and possibly new experiments in secured nuclear laboratories.

In addition to the nuclear implications, new insights into the electronic structure of the plutonium oxide and oxyhydroxide species have been gained, by confronting the results obtained with the two-component relativistic correlated {\dcm}-CCSD method and the more approximate two-step relativistic multireference method SO-CASPT2. Our SO-CASPT2 and {\dcm}-EOM-EE-CCSD results confirm the initial evidence by Feng and Peterson~\cite{actinide-Feng-JPC2017-121-1041} that the ground state of \ce{PuO2} is closed-shell at the relativistic level with a $\mathrm{\Omega = 0_g^+}$ character and provide compelling evidence that it is well separated by about several thousands of \si{\per\cm} from the first open-shell u-state formally assigned as the \ce{PuO2} ground state by La Macchia~{\etal} \cite{actinide-La-Macchia-PCCP2008-10-7278} {\dcm}-EOM-IP-CCSD calculations place the adiabatic ionization energy of \ce{PuO2} at \SI{7.07}{\electronvolt}, in excellent agreement with the up-to-date experimental value. The electronic ground states of \ce{PuO3} and \ce{PuO2(OH)2} exhibit strong multireference characters, and we evidence that multi-reference SO-CASPT2 is a quantitatively appropriate method to compute their enthalpies of formation. The reported values are in good agreement with the available data, making us confident about its predictive capability for other gas-phase heavy element species with complex electronic structures. 

\section*{Acknowledgements}
We gratefully acknowledge financial support from the France-Canada Research Fund, Natural Sciences and Engineering Research Council of Canada (NSERC). This research was supported by the NEEDS environment project (TEMPO, ``ThErmodynamique des Molécules de PlutOnium'') funded in 2017. We acknowledge support by the French government through the Program "Investissement d'avenir" (LABEX CaPPA / ANR-11-LABX-0005-01 and I-SITE ULNE / ANR-16-IDEX-0004 ULNE), as well as by the Ministry of Higher Education and Research, Hauts de France council and European Regional Development Fund (ERDF) through the Contrat de Projets Etat-Region (CPER CLIMIBIO). Furthermore, this work was granted access to the HPC resources of [CINES/IDRIS/TGCC] under the allocation 2016-2019 [x2016081859 and A0010801859, A0030801859, A0050800244] made by GENCI. We also acknowledge  Ma{\l}gorzata Olejniczak for her exploratory work on the benchmarking of relativistic DFT and MS-CASPT2 calculations. We further thank Sidi Souvi and Marc Barrachin for the very fruitful discussions.

\section*{Associated content}
The following file, ms-PuOx-ESI.pdf is available free of charge. It contains:
\begin{itemize}
  \item Mathematical formula used to compute the uncertainties.
  \item Figure illustrating the ANO-RCC basis set convergence to the CBS limit.
  \item Analysis of the spin-free and spin-orbit states of \ce{Pu}, \ce{PuO2}, \ce{PuO3}, and \ce{PuO2(OH)2} and their SO-CASPT2 transition energies, up to \SI{16000}{\per\cm};
  \item Molecular parameters and standard entropies of the \ce{PuO2(g)}, \ce{PuO3(g)}, and \ce{PuO2(OH)2(g)} molecules;
  \item The heat capacity functions of the  \ce{PuO2(g)}, \ce{PuO3(g)}, and \ce{PuO2(OH)2(g)} molecules.
  \item Atomization energies of \ce{PuO2}.
\end{itemize}
The full data set distributed through the Zenodo repository~\cite{actinide-Kervazo-IC2019-Zenodo}.





\newpage
\bibliography{ms-PuOx} 

\end{document}


\clearpage
\tableofcontents
\clearpage
\listoftables
\clearpage
\listoffigures
\clearpage

\section{Details on uncertainties evaluations}
To comply with the standards of the Active Thermochemical Tables developed by Ruscic\cite{}, all standard enthalpies of formation reported in the paper are associated with uncertainties corresponding to the 95\% confidence intervals.

The uncertainties of the derived enthalpies of formation from a given chemical reaction are derived from the formula:

\begin{equation}
2\sigma_m = \left[\sum_i{(2\nu_i\sigma_i})^2\right]^{1/2}
\end{equation}
$2\sigma_i$ are the uncertainties of the standard enthalpies of formation of the species entering the chemical reaction, and $\nu_i$ are the reaction stoichiometric coefficients.

The final standard enthalpies of formation are obtained from a weighted average $\mu$ of the $n$ values $x_i$, ($i=1,\cdots,n$), where each values have an associated uncertainty $\sigma_i$, $i=1,\cdots,n$, and are obtained from the expression~\cite{thermo-Ruscic-JPCRD2005-34-573}:
\begin{equation}
\label{Eq:weightedaverage}
\mu = \frac{\displaystyle\sum_{i=1}^{n}{(x_i/\sigma_i^2})}{\displaystyle\sum_{i=1}^n{(1/{\sigma_i^2})}}
\end{equation}
The variance $s_\mu$ of $\mu$ is then
\begin{equation}
s_\mu = \left(\frac{\displaystyle\sum_{i=1}^n{\left[(x_i-\mu)^2/\sigma_i^2\right]}}{(n-1)\displaystyle\sum_{i=1}^n{1/\sigma_i^2}}\right)^{1/2}
\end{equation}
To bring the final uncertainty closer to the 95\% confidence limits, $s_\mu$ is multiplied by a factor of 2.\cite{thermo-Ruscic-JPCRD2005-34-573}

In our case, the largest uncertainty to the standard enthalpies of formation associated to the available literature data regarding each reaction is that of the plutonium atom with a value of $\pm$3  \si{\kJ\per\mol} (see Table \ref{Tab:litteratureenthalpiesofformation}). All the reactions have the same inherited uncertainty, thus have the same weight in the calculated weighted average of Eq.~\ref{Eq:weightedaverage}. Thus, to also account for the uncertainty of $\sigma_{\mathrm{Pu}}$, the total uncertainty $s_\mu ^{tot} $ (reported in Tables \ref{Tab:RnH_PuO2}, \ref{Tab:RnH_PuO3}, \ref{Tab:RnH_PuO2OH2}) associated to the average over all the considered reaction is calculated as:
\begin{equation}
s_\mu ^{tot} = \sqrt{(2s_{\mu})^2+(2\sigma_{\mathrm{Pu}})^2}
\end{equation}

\clearpage
\section{Basis set convergence to the basis set limit}
\begin{figure}[h]
\begin{subfigure}[t]{0.5\textwidth}
\centering
\includegraphics[width=\linewidth]{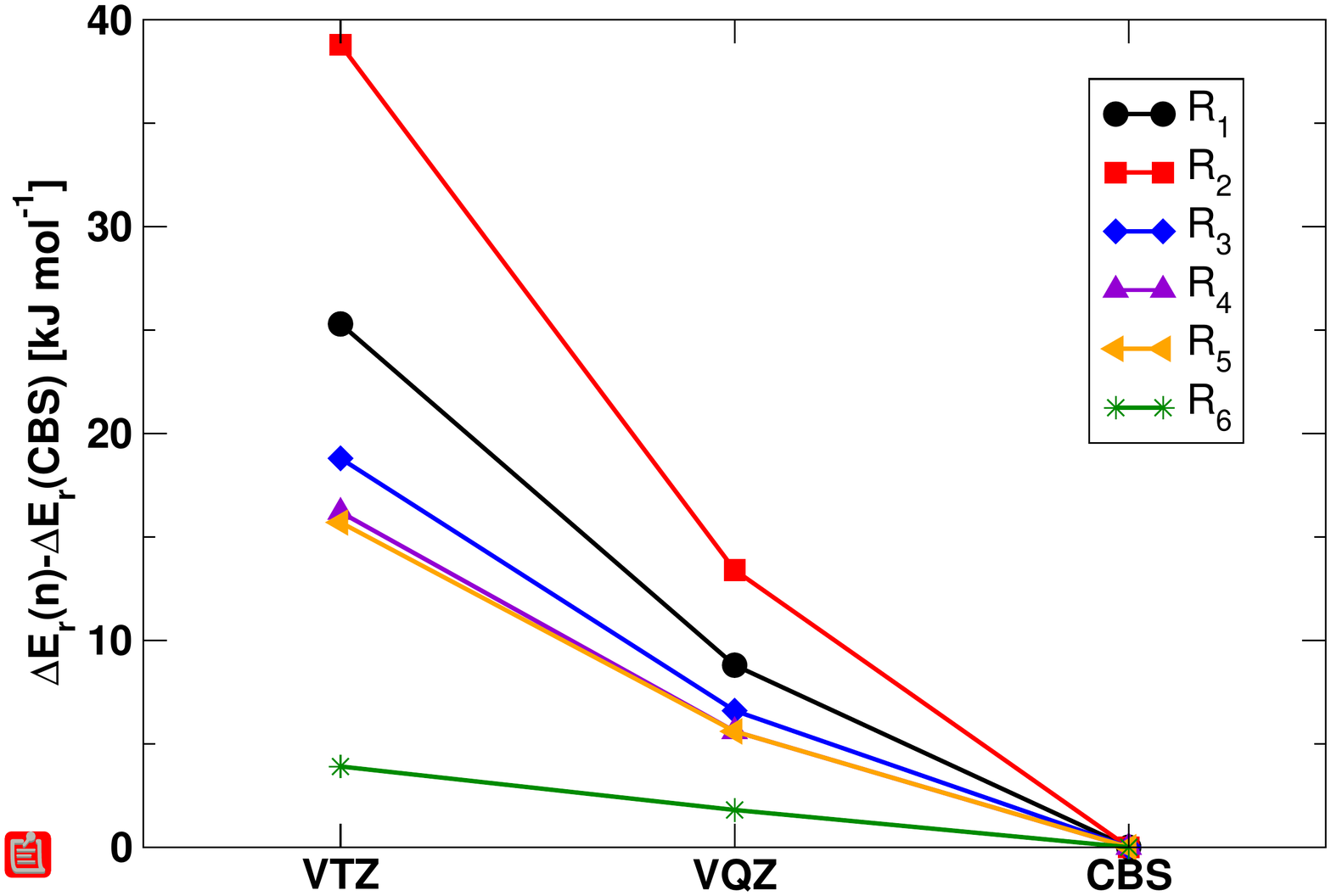}
\caption{\ce{PuO2}}\label{Fig:ANO-RCC-CBS-PuO2}
\end{subfigure}
\begin{subfigure}[t]{0.5\textwidth}
\centering
\includegraphics[width=\linewidth]{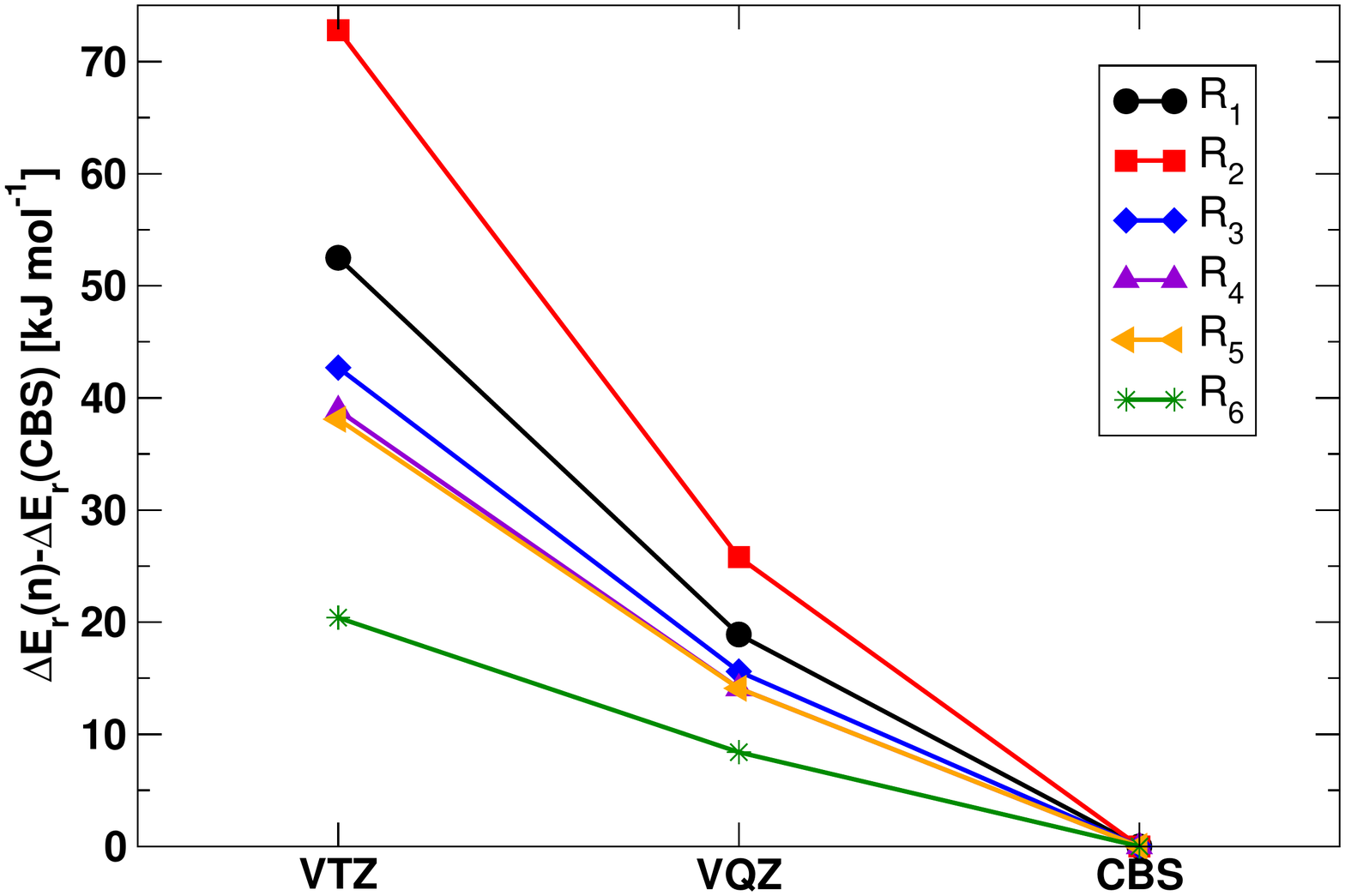}
\caption{\ce{PuO3}}\label{Fig:ANO-RCC-CBS-PuO3}
\end{subfigure}
\begin{subfigure}[t]{0.5\textwidth}
\centering
\includegraphics[width=\linewidth]{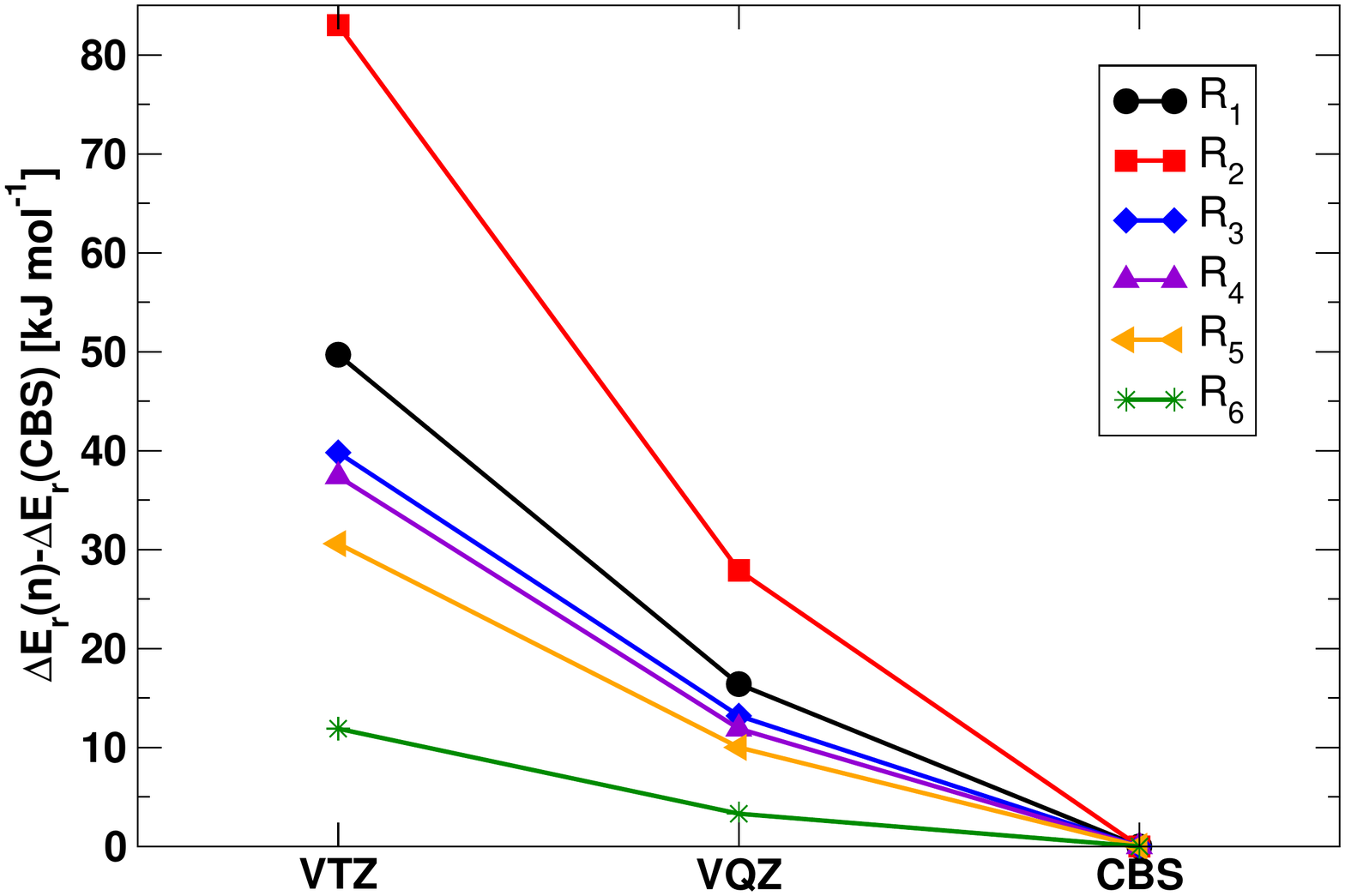}
\caption{\ce{PuO2(OH)2}}\label{Fig:ANO-RCC-CBS-PuO2OH2}
\end{subfigure}
\caption{ANO-RCC basis set convergence to the CBS limit (taken as 0) of the reaction energies at {\Tzero} for the reactions $\mathrm{R_1}$--$\mathrm{R_6}$ leading to the formation of \ce{PuO2} (a), \ce{PuO3} (b) and \ce{PuO2(OH)2} (c).}
\label{Fig:ANO-RCC-CBS}
\end{figure}

All energies have been extrapolated to the complete basis set limit, using Equations~(4) and (5) of the article. For lanthanide complexes, Lu and Peterson~\cite{basis-Lu-JCP2016-145-054111} reported irregular convergence pattern of the ANO-RCC basis sets. However, for Pu complexes, the convergence is smooth, as illustrated by Figure~\ref{Fig:ANO-RCC-CBS}.

\clearpage
\begin{table}[h]
\caption{Standard enthalpies of formation of the molecules in \si{\kJ\per\mol} taken from the Active Thermochemical Tables~\cite{thermo-Ruscic-2019-ATcT-1.122e} and from Ref.~\cite{actinide-Lemire-ChemThermo-2001} for Pu; Zero-Point Vibrational energy and enthalpy increment \state[pre=]{H}(\Tref)-\state[pre=]{H}(\SI{0}{\kelvin}) computed at the B3LYP level with aug-cc-pVTZ basis sets.}
\begin{center}
\begin{tabular}{lS[table-format=4.4(4)]SS}
\toprule
Species  &  {\enthalpy*(f){}(\Tref)} & {ZPE} & {\state[pre=]{H}(\Tref)-\state[pre=]{H}(\SI{0}{\kelvin})}\\
\midrule
O ($^3P$)  &  249.229 \pm 0.0021  &  0 & 6.73 \\
H ($^1S$)   &  217.998\pm0.000 & 0 & 6.20 \\
Pu ($^5F$)  &  349.00\pm3.0  	& 0& 6.20 \\
\ce{H2}   &  0.0  &26.44& 8.68\\
\ce{H2O}  &  -241.835\pm0.027  &  53.10 & 9.93\\
\ce{H2O2}  &  -135.464\pm0.063 & 69.33 & 10.96\\
\ce{O2}  &  0.0&  9.72 & 8.68 \\
\ce{OH}  &  37.489\pm0.027  & 22.08 & 8.68\\
\ce{PuO2} & & 11.50 & 12.35\\
\ce{PuO3} && 16.32& 15.34\\
\ce{PuO2(OH)2} & &  80.47 & 21.33\\
\ce{PuO2+} && 14.68 & 11.41\\
\ce{PuO3+} && 18.07 & 14.73\\
\bottomrule
\end{tabular}
\end{center}
\label{Tab:litteratureenthalpiesofformation}
\end{table}%

\newpage

\begin{table} 
\caption{Fine structure transition energies ($\Delta E$ in \si{\per\cm}) of atomic Pu computed at the SO-CASPT2 level with the ANO-RCC-TZVP basis set, and analysis of the various $J$-states in terms of the dominating $LS$ terms.}
\label{Tab:All-levelsPu}
\begin{center}
\begin{tabular}{*2Sl}
\toprule
 {$J$-value} &	{$\Delta E$}  &  Weight of $LS$ states	\\
\midrule
0	& 0	&	54\%	 $\mathrm{^7F}$,	38\%	 $\mathrm{^5D}$,	7\%	 $\mathrm{^3P}$ \\	
1	& 1806	 &	64\%	 $\mathrm{^7F}$,	32\%	 $\mathrm{^5D}$			\\	
2	& 4208	 &	82\%	 $\mathrm{^7F}$,	17\%	 $\mathrm{^5D}$			\\	
3	& 6353	 &	88\%	 $\mathrm{^7F}$,	10\%	 $\mathrm{^5D}$			\\	
4	& 8091	 &	93\%	 $\mathrm{^7F}$							\\
5	& 9552	 &	90\%	 $\mathrm{^7F}$							\\
6	& 11010	&	85\%	 $\mathrm{^7F}$,	10\%	 $\mathrm{^5G}$		\\		
0	& 12499	&	44\%	 $\mathrm{^7F}$,		21\%	 $\mathrm{^5D}$,	8\%	 $\mathrm{^3P}$ \\
1	& 13612	&	66\%	 $\mathrm{^9H}$, 9\%	 $\mathrm{^7F}$,			7\%	 $\mathrm{^7G}$ \\
2	& 14399	&	82\%	 $\mathrm{^9H}$,	14\%	 $\mathrm{^7G}$				 \\
1	& 14845	&	38\%	 $\mathrm{^5D}$, 20\%	 $\mathrm{^7F}$, 9\%	 $\mathrm{^9G}$		\\		
3	& 15551	&	89\%	 $\mathrm{^9H}$,	8\%	 $\mathrm{^7G}$				\\
4	& 17026	&	86\%	 $\mathrm{^9H}$,	6\%	 $\mathrm{^7G}$				\\
1	& 17032	&	66\%	 $\mathrm{^9D}$,	29\%	 $\mathrm{^7P}$				\\ 
\bottomrule
\end{tabular}
\end{center}
\end{table} 

\clearpage
\begin{footnotesize}
\begin{longtable}{*{2}{l}S[table-format=5.0]S[table-format=1.2]}
\caption{Vertical transition energies ($\Delta E$ in \si{\per\cm}, up to \SI{16 000}{\per\cm}) of \ce{PuO3} at the SF-CASPT2 and SO-CASPT2 levels obtained with the ANO-RCC-TZVP. Weights of the reference wavefunction in the CASPT2 perturbation treatment are given for the SF states. Orbitals 1a$_1$, 2a$_1$/$\mathrm{\pi_u}$, 3a$_1$ being always doubly occupied, they do not appear in the following table.}\\
\label{Tab:All-levelsPuO3}\\
\toprule
State &	Character	:	\%	[orbital(number	of	electron)]	& {$\Delta E$} & {Ref. weight}\\
\midrule
\endfirsthead
\multicolumn{4}{c}%
{\tablename\ \thetable\ -- \textit{Continued from previous page}} \\
\midrule
State &	Character	:	\%	[orbital(number	of	electron)]	&{$\Delta E$} 	& {Ref. weight}\\
\midrule
\endhead
\bottomrule
\endfoot
\multicolumn{4}{c}{SF-CASPT2 states}\\				
$\mathrm{(1)^3B_2}$	  	&	  	36\%  	[4a$_1$/p$^{(O)}$(2),	1b$_1$/$\mathrm{\phi_u}$(1),	1b$_2$/p$^{(O)}$($\overline{1}$),	2b$_2$/$\mathrm{\sigma_u}$(2),	3b$_2$/$\mathrm{\delta_u}$(1),	1a$_2$/$\mathrm{\delta_u}$(1)]  				&	  	0	  	& 0.72\\
		&	21\%	[4a$_1$/p$^{(O)}$(2),	1b$_1$/$\mathrm{\phi_u}$(1),	1b$_2$/p$^{(O)}$(1),	2b$_2$/$\mathrm{\sigma_u}$(2),	3b$_2$/$\mathrm{\delta_u}$($\overline{1}$),	1a$_2$/$\mathrm{\delta_u}$(1)]  		&		\\
		&	22\%	[4a$_1$/p$^{(O)}$(2),1b$_1$/$\mathrm{\phi_u}$(1),1b$_2$/p$^{(O)}$(2),2b$_2$/$\mathrm{\sigma_u}$(2),1a$_2$/$\mathrm{\delta_u}$(1)]  												&		\\
	  	$\mathrm{(1)^3A_2}$	  	&	43\%	[4a$_1$/p$^{(O)}$(2),	5a$_1$/$\mathrm{\phi_u}$(1),1b$_2$/p$^{(O)}$($\overline{1}$),2b$_2$/$\mathrm{\sigma_u}$(2),3b$_2$/$\mathrm{\delta_u}$(1),1a$_2$/$\mathrm{\delta_u}$(1)]  										&	  	3161	  	& 0.72\\
		&	23\%	[4a$_1$/p$^{(O)}$(2),	5a$_1$/$\mathrm{\phi_u}$(1),1b$_2$/p$^{(O)}$(1),2b$_2$/$\mathrm{\sigma_u}$(2),3b$_2$/$\mathrm{\delta_u}$($\overline{1}$),1a$_2$/$\mathrm{\delta_u}$(1)]  											&		\\
		&	17\%	[4a$_1$/p$^{(O)}$(2),	5a$_1$/$\mathrm{\phi_u}$(1),1b$_2$/p$^{(O)}$(2),2b$_2$/$\mathrm{\sigma_u}$(2),1a$_2$/$\mathrm{\delta_u}$(1)]  											&		\\
	  	$\mathrm{(1)^3B_1}$	  	&	  	38\%  	[4a$_1$/p$^{(O)}$(2),5a$_1$/$\mathrm{\phi_u}$(1),1b$_1$/$\mathrm{\phi_u}$(1),1b$_2$/p$^{(O)}$($\overline{1}$),2b$_2$/$\mathrm{\sigma_u}$(2),3b$_2$/$\mathrm{\delta_u}$(1)]	  											&	  	4663	  	& 0.72\\
\bottomrule
\end{longtable} 
\end{footnotesize}

\clearpage
\begin{longtable}{*{2}{l}S[table-format=5.0]S[table-format=1.2]} 
\caption{Vertical transition energies ($\Delta E$ in \si{\per\cm}, up to \SI{16 000}{\per\cm}) of \ce{PuO2(OH)2} at the SF-CASPT2 and SO-CASPT2 levels obtained with the ANO-RCC-TZVP basis sets. Weights of the reference wavefunction in the CASPT2 perturbation treatment are given for the SF states.}\\
\label{Tab:All-levelsPuO2OH2}\\
\toprule
State &	Character	:	\%	[orbital(number	of	electron)]	&	{$\Delta E$} & {Ref. weight}	\\
\endfirsthead
\multicolumn{4}{c}%
{\tablename\ \thetable\ -- \textit{Continued from previous page}} \\
\toprule
State &	Character	:	\%	[orbital(number	of	electron)]	&	{$\Delta E$} & {Ref. weight}	\\
\midrule
\endhead
\bottomrule
\endfoot
\midrule	
\multicolumn{4}{c}{SF-CASPT2 states}\\	
	$\mathrm{(1)^3B}$	&	49\% 2a/$\mathrm{\phi_u}$(1), 1b/$\mathrm{\delta_u}$(1)  	&	0	& 0.67\\
		&	48\% 1a/$\mathrm{\delta_u}$(1),   2b/$\mathrm{\phi_u}$(1)	&		\\
	$\mathrm{(1)^3A}$	&	  41\%  1a/$\mathrm{\delta_u}$(1),           2a/$\mathrm{\phi_u}$(1)	&	889	& 0.67\\
		&	  59\% 1b/$\mathrm{\delta_u}$(1),             2b/$\mathrm{\phi_u}$(1)	&		\\
	$\mathrm{(2)^3B}$	&	75\% 1a/$\mathrm{\delta_u}$(1),   1b/$\mathrm{\delta_u}$(1)	&	2136	 & 0.67\\
		&	20\% 2a/$\mathrm{\phi_u}$(1), 2b/$\mathrm{\phi_u}$(1)  	&		\\
	$\mathrm{(2)^3A}$	&	  59\% 1a/$\mathrm{\delta_u}$(1),             2a/$\mathrm{\phi_u}$(1)	&	4698	 & 0.67\\
\end{longtable} 

\clearpage
\begin{footnotesize}
\begin{longtable}{*{2}{l}S[table-format=5.0]S[table-format=1.2]} 
\caption{Vertical transition energies ($\Delta E$ in \si{\per\cm}, up to \SI{16 000}{\per\cm}) of  \ce{PuO2} ((Distance \ce{Pu-O} = \SI{1.808}{\angstrom}) at SF-CASPT2 and SO-CASPT2 levels obtained with the ANO-RCC-TZVP basis sets. Weights of the reference wavefunction in the CASPT2 perturbation treatment are given for the SF states. The 1a$_g$, 1b$_{1u}$/$\sigma$$\mathrm{_u}$ and 1b$_{3u}$/2p orbitals being always doubly occupied, they do not appear in the table.}\\
\label{Tab:All-levelsPuO2}\\
\toprule
 State &	Character	:	\%	[orbital(number	of	electron)]	&	{$\Delta E$}  & {Ref. weight}\\
\endfirsthead
\multicolumn{4}{c}%
{\tablename\ \thetable\ -- \textit{Continued from previous page}} \\
\midrule
State &	Character	:	\%	[orbital(number	of	electron)]	&	{$\Delta E$}	 & {Ref. weight}\\
\midrule
\endhead
\bottomrule
\endfoot
\multicolumn{4}{c}{SF-CASPT2 states}\\
$\mathrm{(1)^{5}A_g}$	 	&	 	84\% 	[1a$\mathrm{_u}$/$\delta$$\mathrm{_u}$(1),	2b$_{2u}$/$\Phi$$\mathrm{_u}$(1),	2b$_{1u}$/$\delta$$\mathrm{_u}$(1),	2b$_{3u}$/$\Phi$$\mathrm{_u}$(1)]	  		&	 	0	& 0.78 \\
		 	$\mathrm{(2)^{5}A_g}$	 	&	 	45\% 	[1a$\mathrm{_u}$/$\delta$$\mathrm{_u}$(1),	3b$_{2u}$/$\pi$$\mathrm{_u}$(1),	2b$_{1u}$/$\delta$$\mathrm{_u}$(1),	2b$_{3u}$/$\Phi$$\mathrm{_u}$(1)]	 		&	 	4798	 & 0.78\\
			&	45\%	[1a$\mathrm{_u}$/$\delta$$\mathrm{_u}$(1),	2b$_{2u}$/$\Phi$$\mathrm{_u}$(1),	2b$_{1u}$/$\delta$$\mathrm{_u}$3(1),	3b$_{3u}$/$\pi$$\mathrm{_u}$(1)]	 		&	\\
		 	$\mathrm{(1)^{5}B_{3u}}$	 	&	 	85\% 	[3a$_g$(1),	1a$\mathrm{_u}$/$\delta$$\mathrm{_u}$(1),	2b$_{1u}$/$\delta$$\mathrm{_u}$(1),	2b$_{3u}$/$\Phi$$\mathrm{_u}$(1)] 			&	 	5421	 & 0.78\\
		 	$\mathrm{(1)^{5}B_{2u}}$	 	&	 	85\% 	[3a$_g$(1),	1a$\mathrm{_u}$/$\delta$$\mathrm{_u}$(1),	2b$_{1u}$/$\delta$$\mathrm{_u}$(1),	2b$_{3u}$/$\Phi$$\mathrm{_u}$(1)] 			&	 	5421	 \\
		 	$\mathrm{(1)^{5}B_{3g}}$	 	&	44\%	[2b$_{2u}$/$\Phi$$\mathrm{_u}$(1),	3b$_{2u}$/$\pi$$\mathrm{_u}$1(1),	2b$_{1u}$/$\delta$$\mathrm{_u}$(1),	2b$_{3u}$/$\Phi$$\mathrm{_u}$(1)] 			&	 	6019	   & 0.78\\
			&	44\%	[1a$\mathrm{_u}$/$\delta$$\mathrm{_u}$(1),	2b$_{2u}$/$\Phi$$\mathrm{_u}$(1),	2b$_{3u}$/$\Phi$$\mathrm{_u}$(1),	3b$_{3u}$/$\pi$$\mathrm{_u}$(1)]	 		&	\\
\end{longtable} 
\end{footnotesize}

\newpage

\begin{table}[t] 
\caption{Number of electronic states at MS-CASPT2 level of calculations.}
\label{Tab:numberstatePT2}
\begin{tabular}{l*{15}{r}}
\toprule
   Multiplicity & Pu & \multicolumn{8}{c}{\ce{PuO2}} &  \multicolumn{4}{c}{\ce{PuO3}}&  \multicolumn{2}{c}{\ce{PuO2(OH)2}}\\
\cmidrule(r){1-1}\cmidrule(r){2-2}\cmidrule(r){3-10}\cmidrule(r){11-14}\cmidrule(r){15-16}
 symmetry & {$\mathrm{A_g}$} & {$\mathrm{A_g}$} & {$\mathrm{B_{2g}}$} & {$\mathrm{B_{1g}}$} & {$\mathrm{B_{3g}}$} & {$\mathrm{A_u}$} & {$\mathrm{B_{2u}}$} &  {$\mathrm{B_{1u}}$} &  {$\mathrm{B_{3u}}$} & {$\mathrm{A_1}$} & {$\mathrm{B_1}$} & {$\mathrm{B_2}$} & {$\mathrm{A_2}$} & {A} & {B} \\ 
 \midrule
  Singlet & 0 & 2 & 0 & 2 & 0 & 3 & 3 & 3 & 3 & 10 & 9 & 7 & 7 & 7 & 4 \\
   Triplet &  27 & 2 & 3 & 3 & 3 & 6 & 7 & 6 & 7 & 16 & 15 & 13 & 14 & 2 & 4 \\
   Quintet & 113 & 4 & 3 & 2 & 3 & 4 & 4 & 4 & 4 & 8 & 7 & 6 & 7 & 0 & 0 \\
   Septet & 103 &  0 & 0 & 0 & 0 & 0 & 0 & 0 & 0 & 0 & 0 & 0 & 0 & 0 & 0\\   
   Nonet & 35 & 0& 0& 0& 0& 0& 0& 0& 0& 0& 0& 0&0 &0 &0 \\
\bottomrule
\end{tabular}    
\end{table}

\begin{table}[h]
\begin{center}
\caption{The molecular parameters of \ce{PuO2(g)} obtained at the CBS {\dcm}-CCSD(T) level. All electronic states (and their degeneracy in brackets) are not written here, but any levels listed in Table~6 (CBS-{\dcm}-EOM-CCSD values) are taken into account  in the calculation of thermodynamic functions.}
\label{Tab:paramPuO2}
\begin{tabular}{lc} 
\toprule
Parameter&Data \\ \midrule 
Symmetry -- sym. number& $D_{\infty h}$ -- 2 \\
Moment of inertia (g.cm$^{-2})$& 1.75$\times$10$^{-38}$ \\
Vibrational frequencies (cm$^{-1}$)&170(2), 791, 840 \\
Electronic states (cm$^{-1}$)&0(1), 2457(2), 5663(2), 7134(1), 9073(2), 9502(2), etc\\
\bottomrule
 \end{tabular}
\end{center}
\end{table} 

\begin{table}[h]
\begin{center}
\caption{The molecular parameters of \ce{PuO3(g)}.  Electronic states are not degenerate, all levels are not written here, but any states listed in Table~\ref{Tab:All-levelsPuO3} are taken into account in the calculation of thermodynamic functions.}
\label{Tab:paramPuO3}
\begin{tabular}{lc} \toprule
Parameter&Data \\ \midrule 
Symmetry -- sym. number& $C_{2v}$ -- 2 \\
Moment of inertia (g$^{3}$.cm$^{-6})$& 4.14$\times$10$^{-114}$ \\
Vibrational frequencies (cm$^{-1}$)&195, 220, 238, 335, 797, 893 \\
Electronic states (cm$^{-1}$)&0, 1235, 1783, 3777, 5660, 6658, 7062, etc\\
\bottomrule
\end{tabular}
\end{center}
\end{table} 

\begin{table}[h]
\begin{center}
\caption{The molecular parameters of \ce{PuO2(OH)2(g)}.   Electronic states are not degenerate, all levels are not written here, but any states listed in Table~\ref{Tab:All-levelsPuO2OH2} are taken into account in the calculation of thermodynamic functions.}
\label{Tab:paramPuH2O4}
\begin{tabular}{lc} \toprule
Parameter&Data \\ \midrule 
Symmetry -- sym. number& $C_{2}$ -- 2 \\
Moment of inertia (g$^{3}$.cm$^{-6})$& 2.044$\times$10$^{-113}$ \\
Vibrational frequencies (cm$^{-1}$)&103, 185, 225, 240, 270, 386, 415, 491 \\
& 504, 547, 559, 857, 960, 3855, 3857 \\
Electronic states (cm$^{-1}$)&0, 323, 2222, 3757, 4158, 7268, 7575, etc\\
\bottomrule
 \end{tabular}
\end{center}
\end{table} 

\begin{table}[h]
\begin{center}
\caption{Standard entropies at \SI{298.15}{\kelvin} (in \si{\joule\per\kelvin\per\mol}) with detailed contributions. $^{a}$Moments of inertia have been calculated from geometries got in Ref.~\cite{actinide-Ebbinghaus-TR-2002} within Gaussian09 package. $^{b}$Two internal rotations have been treated by author with hindered rotor instead harmonic oscillator.}
\label{Tab:paramPuOx}
\begin{tabular}{l*7S} 
\toprule
Species&\multicolumn{2}{c}{\ce{PuO2(g)}}&\multicolumn{3}{c}{\ce{PuO3(g)}}&\multicolumn{2}{c}{\ce{PuO2(OH)2(g)}} \\ 
\cmidrule(r){2-3}\cmidrule(r){4-6}\cmidrule(r){7-8}
Source&{this work}&{Ref.~\cite{actinide-Konings-JPCRD2014-43-013101}}&
{this work}&{Ref.~\cite{actinide-Konings-JPCRD2014-43-013101}}&{Ref.~\cite{actinide-Ebbinghaus-TR-2002}}&
{this work}&{Ref.~\cite{actinide-Ebbinghaus-TR-2002}}\\ 
\midrule
Translational&178.7&178.7&179.4&179.4&179.4&180.2&180.2 \\
Rotational&62.1&62.4&99.8&100.4&97.3$^{a}$&106.2&{105.9$^{a}$ (+35.9$^{b}$)}\\
Vibrational&22.0&29.2&31.6&33.9&31.6&65.3&48.0 \\
Electronic&0.0&8.5&0.2&5.8&7.7&3.8&7.7\\ \midrule
\state[pre=,subscript-right={\SI{298}{\kelvin}}]{S}&262.9&278.7&311.0&319.4&316.1&355.7&377.8 \\
\bottomrule
\end{tabular}
\end{center}
\end{table} 

\clearpage
\begin{table}[h]
\begin{center}
\caption{\SI{0}{\kelvin} atomization energies ($\mathrm{AE_0}$, in \si{\kJ\per\mol}) of \ce{PuO2}, corresponding to the opposite of Reaction R2 at various levels of theory}
\label{Tab:R2}
\begin{tabular}{lS}
\toprule
Method & {$\mathrm{AE_0}$}\\
\midrule
{SO-B3LYP} 		& 1157.9\\
{SO-UCCSD} 		& 1172.6\\
{SO-UCCSD(T)} 	& 1255.7\\
{SO-CASPT2} 		& 1254.6\\
{SO-CCSD(T)} [\cite{actinide-Feng-JCP2017-147-084108}] & 1278.6\\
\bottomrule
\end{tabular}
\end{center}
\end{table} 

\clearpage
\begin{small}
\begin{longtable}{ll*{6}{S[round-mode=places,round-precision=1]}}
\caption{Summary of contributions to the composite calculation of the reaction enthalpies leading to \ce{PuO2} in \si{\kJ\per\mol}, and derived standard enthalpy of formation with its uncertainty. The final average values includes all reactions, except for the {\dcm}-CCSD and {\dcm}-CCSD(T) values which are averaged over reactions $\mathrm{R_3}$--$\mathrm{R_6}$. \label{Tab:RnH_PuO2}}\\
\toprule
{Rn.} & {Method} & {$\Delta E_{SF}$} & {$\Delta \mathrm{ZPE}$} & {$\Delta \mathrm{SO}$} & {\enthalpy*(r){}{}({\Tzero})} & {\enthalpy*(r){}{}({\Tref})} & {\enthalpy*(f){}{}({\Tref})} \\
\midrule
\endfirsthead
\multicolumn{7}{c}%
{\tablename\ \thetable\ -- \textit{Continued from previous page}} \\
\toprule
{Rn.} & {Method} & {$\Delta E_{SF}$} & {$\Delta \mathrm{ZPE}$} & {$\Delta \mathrm{SO}$} & {\enthalpy*(r){}{}({\Tzero})} & {\enthalpy*(r){}{}({\Tref})} & {\enthalpy*(f){}{}({\Tref})} \\
\midrule
\endhead
\bottomrule
\endfoot
R$_{1}$ 
& SO-B3LYP &-680.0 & {\multirow{4}{*}{1.8}} & {\multirow{4}{*}{37.4}} & -640.8 & -643.3 & -294.3 \pm3.0\\
& SO-UCCSD & -750.8 &  &  & -711.6 & -714.1 & -365.1\pm3.0\\
& SO-UCCSD(T) & -795.4 & &  & -756.2 & -758.8 & -409.8\pm3.0\\
& SO-CASPT2 & -737.29 &  & & -698.1 & -700.6 & -351.6\pm3.0\\
\midrule
R$_{2}$ 
& SO-B3LYP & -1195.3 & {\multirow{4}{*}{11.5}} & {\multirow{4}{*}{37.4}} & -1146.4 & -1153.7 & -306.2\pm3.0\\
& SO-UCCSD & -1210.01 &  &  & -1161.1 & -1168.4 & -320.9\pm3.0\\
& SO-UCCSD(T) & -1293.12 & &  & -1244.2 & -1251.5 & -404.0\pm3.0\\
& SO-CASPT2 & -1292.0 &  & & -1243.1 & -1250.4 & -402.9\pm3.0\\
\end{longtable}
\end{small}

\clearpage
\begin{small}
\begin{longtable}{ll*{6}{S[round-mode=places,round-precision=1]}}
\caption{Summary of contributions to the composite calculation of the reaction enthalpies leading to \ce{PuO3} in \si{\kJ\per\mol}, and derived standard enthalpy of formation with its uncertainty. The final average values includes all reactions. \label{Tab:RnH_PuO3}}\\
\toprule
{Rn.} & {Method} & {$\Delta E_{SF}$} & {$\Delta \mathrm{ZPE}$} & {$\Delta \mathrm{SO}$} & {\enthalpy*(r){}{}({\Tzero})} & {\enthalpy*(r){}{}({\Tref})} & {\enthalpy*(f){}{}({\Tref})} \\
\midrule
\endfirsthead
\multicolumn{7}{c}%
{\tablename\ \thetable\ -- \textit{Continued from previous page}} \\
\toprule
{Rn.} & {Method} & {$\Delta E_{SF}$} & {$\Delta \mathrm{ZPE}$} & {$\Delta \mathrm{SO}$} & {\enthalpy*(r){}{}({\Tzero})} & {\enthalpy*(r){}{}({\Tref})} & {\enthalpy*(f){}{}({\Tref})} \\
\midrule
\endhead
\bottomrule
\endfoot
R$_{1}$ 
& SO-B3LYP & -781.6& {\multirow{2}{*}{1.7}} & {\multirow{2}{*}{59.0}} &  -720.9 & -724.8 & -375.8\pm3.0\\
& SO-CASPT2 & -919.7 &  & & -859.0 & -862.8 & -513.8\pm3.0\\
\midrule

R$_{2}$ 
& SO-B3LYP & -1554.6& {\multirow{2}{*}{16.3}} & {\multirow{2}{*}{59.0}} &  -1479.3 &-1490.3 & -393.6\pm3.0\\
& SO-CASPT2 & -1697.7 &  & & -1622.5 & -1633.5 & -536.8\pm3.0\\
\midrule

R$_{3}$ 
& SO-B3LYP & -571.7& {\multirow{2}{*}{-48.0}} & {\multirow{2}{*}{59.0}} &  -560.7 & -555.0 & -409.2\pm3.0\\
& SO-CASPT2 & -694.4 &  & & -683.4 & -677.7 & -531.9\pm3.0\\
\midrule

R$_{4}$ 
& SO-B3LYP & -198.5& {\multirow{2}{*}{-49.9}} & {\multirow{2}{*}{59.0}} & -189.5 & -187.8 & -380.3\pm3.0\\
& SO-CASPT2 & -373.3 &  & & -364.3 & -362.6	 & -555.1\pm3.0\\
\end{longtable}
\end{small}

\clearpage
\begin{small}
\begin{longtable}{ll*{6}{S[round-mode=places,round-precision=1]}}
\caption{Summary of contributions to the composite calculation of the reaction enthalpies leading to \ce{PuO2(OH)2} in \si{\kJ\per\mol}, and derived standard enthalpy of formation with its uncertainty. The final average values includes all reactions. \label{Tab:RnH_PuO2OH2}}\\
\toprule
{Rn.} & {Method} & {$\Delta E_{SF}$} & {$\Delta \mathrm{ZPE}$} & {$\Delta \mathrm{SO}$} & {\enthalpy*(r){}{}({\Tzero})} & {\enthalpy*(r){}{}({\Tref})} & {\enthalpy*(f){}{}({\Tref})} \\
\midrule
\endfirsthead
\multicolumn{7}{c}%
{\tablename\ \thetable\ -- \textit{Continued from previous page}} \\
\toprule
{Rn.} & {Method} & {$\Delta E_{SF}$} & {$\Delta \mathrm{ZPE}$} & {$\Delta \mathrm{SO}$} & {\enthalpy*(r){}{}({\Tzero})} & {\enthalpy*(r){}{}({\Tref})} & {\enthalpy*(f){}{}({\Tref})} \\
\midrule
\endhead
\bottomrule
\endfoot
R$_{1}$ 
& SO-B3LYP & -1184.2& {\multirow{2}{*}{34.6}} & {\multirow{2}{*}{46.9}} &  -1102.6 & -1113.6 & -764.6\pm3.0\\
& SO-CASPT2 &-1389.99 &  & & -1308.5 & -1319.4 & -970.4\pm3.0\\
\midrule

R$_{2}$ 
& SO-B3LYP & -2675.5& {\multirow{2}{*}{80.5}} & {\multirow{2}{*}{46.9}} & -2548.1 & -2572.2 & -790.3\pm3.0\\
& SO-CASPT2 &-2875.8 &  & & -2748.3 & -2772.5 & -990.6\pm3.0\\
\midrule

R$_{3}$ 
& SO-B3LYP & -1242.1& {\multirow{2}{*}{-21.3}} & {\multirow{2}{*}{46.9}} & -1216.5 & -1214.4 & -788.1\pm3.0\\
& SO-CASPT2 &-1420.0 &  & & -1394.4 & -1392.3 & -966.0\pm3.0\\
\midrule

R$_{4}$ 
& SO-B3LYP & -867.5& {\multirow{2}{*}{-7.8}} & {\multirow{2}{*}{46.9}} & -828.4 & -835.6 & -772.6\pm3.0\\
& SO-CASPT2 &-1109.6 &  & & -1070.5 & -1077.6& -1014.7\pm3.0\\
\end{longtable}
\end{small}

\clearpage

\begin{figure}
\centering
\includegraphics[width=12cm]{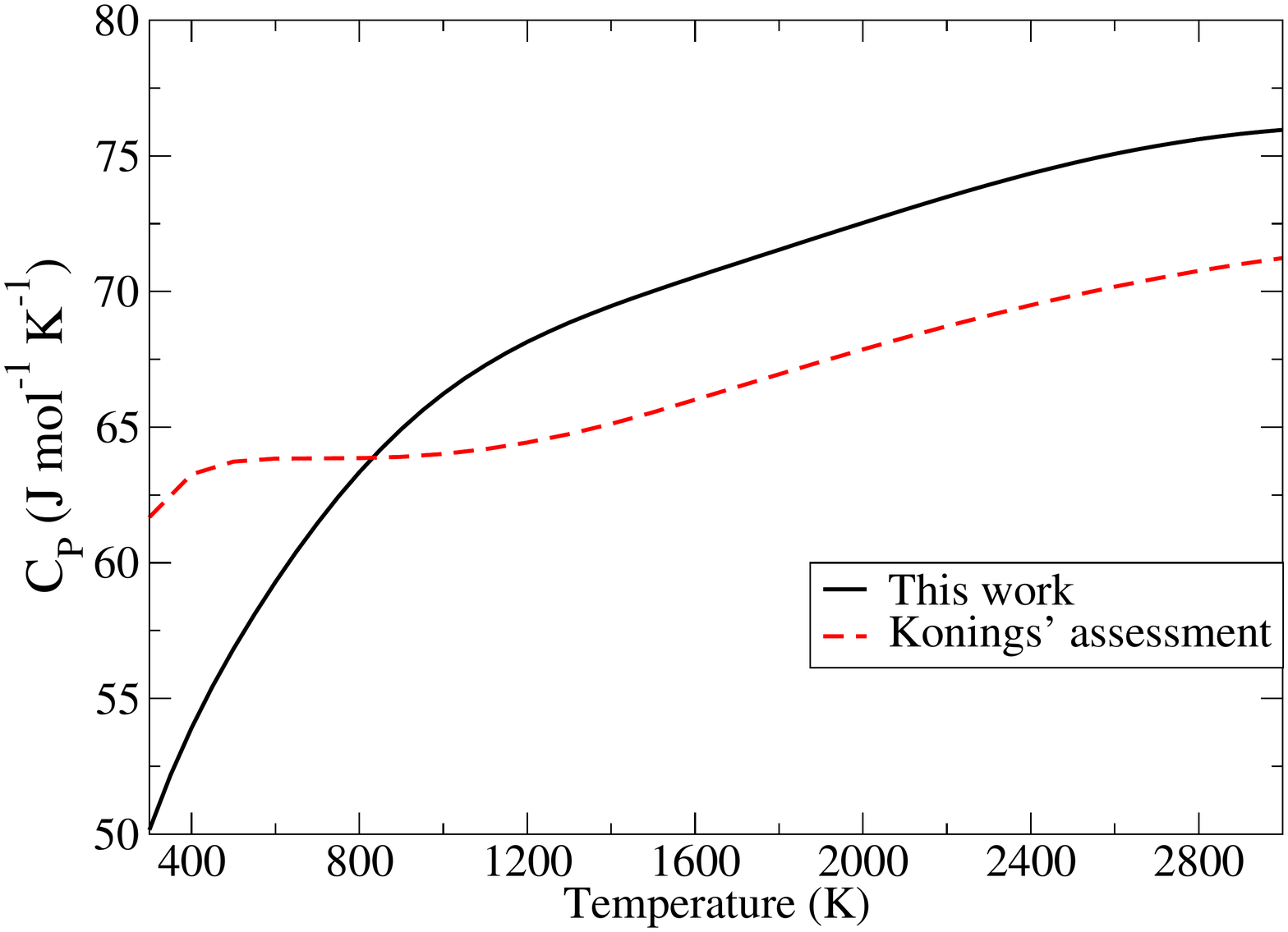}
\caption{Heat capacity of \ce{PuO2(g)}. Konings' assessment refers to Ref.~\cite{actinide-Konings-JPCRD2014-43-013101}.}
\label{Fig:molPuO2}
\end{figure}

\begin{figure}
\centering
\includegraphics[width=12cm]{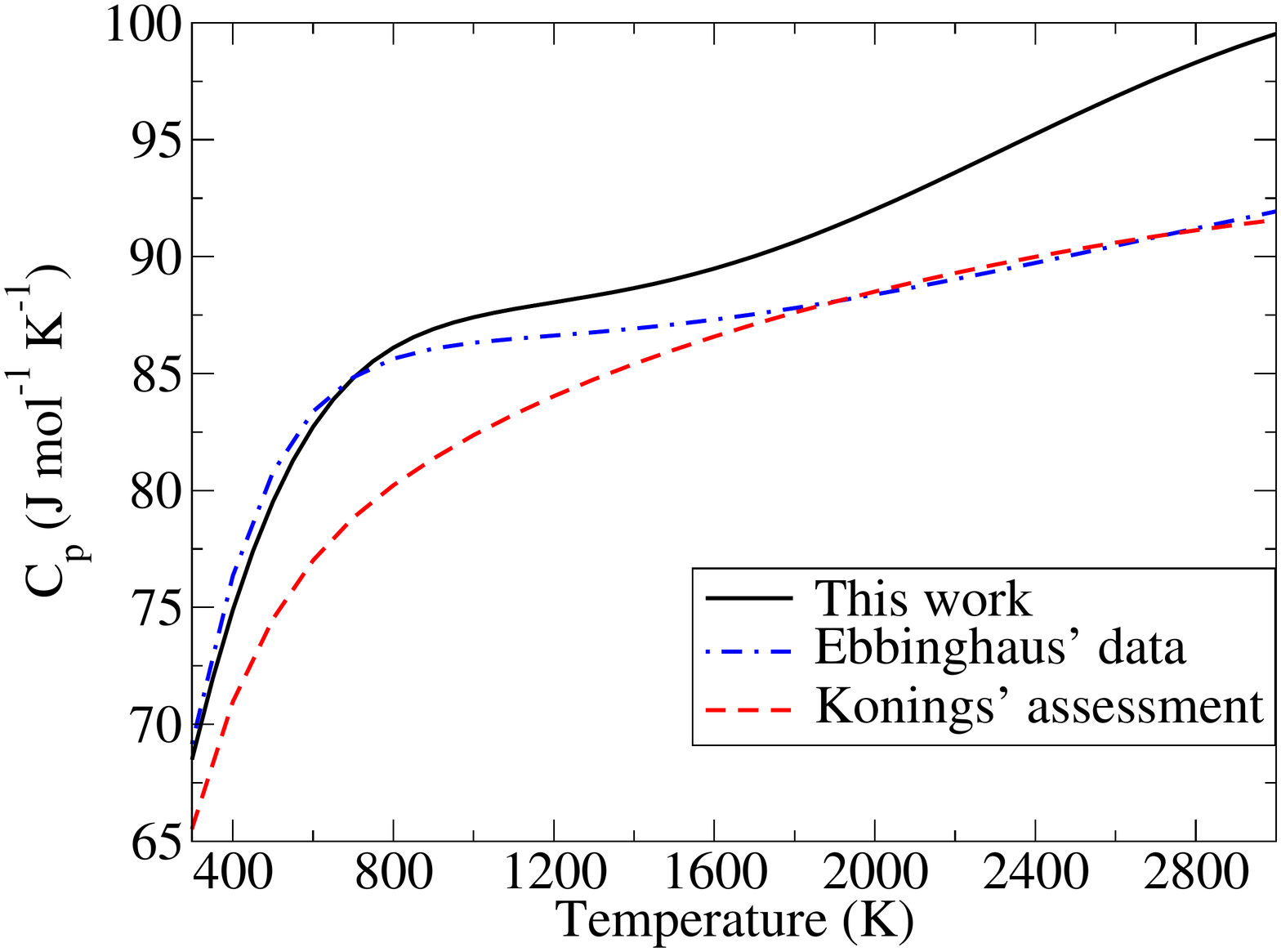}
\caption{Heat capacity of \ce{PuO3(g)}. Konings' assessment refers to Ref.~\cite{actinide-Konings-JPCRD2014-43-013101} and Ebbinghaus' data to Ref. \cite{actinide-Ebbinghaus-TR-2002}}
\label{Fig:molPuO3}
\end{figure}

\begin{figure}
\centering
\includegraphics[width=12cm]{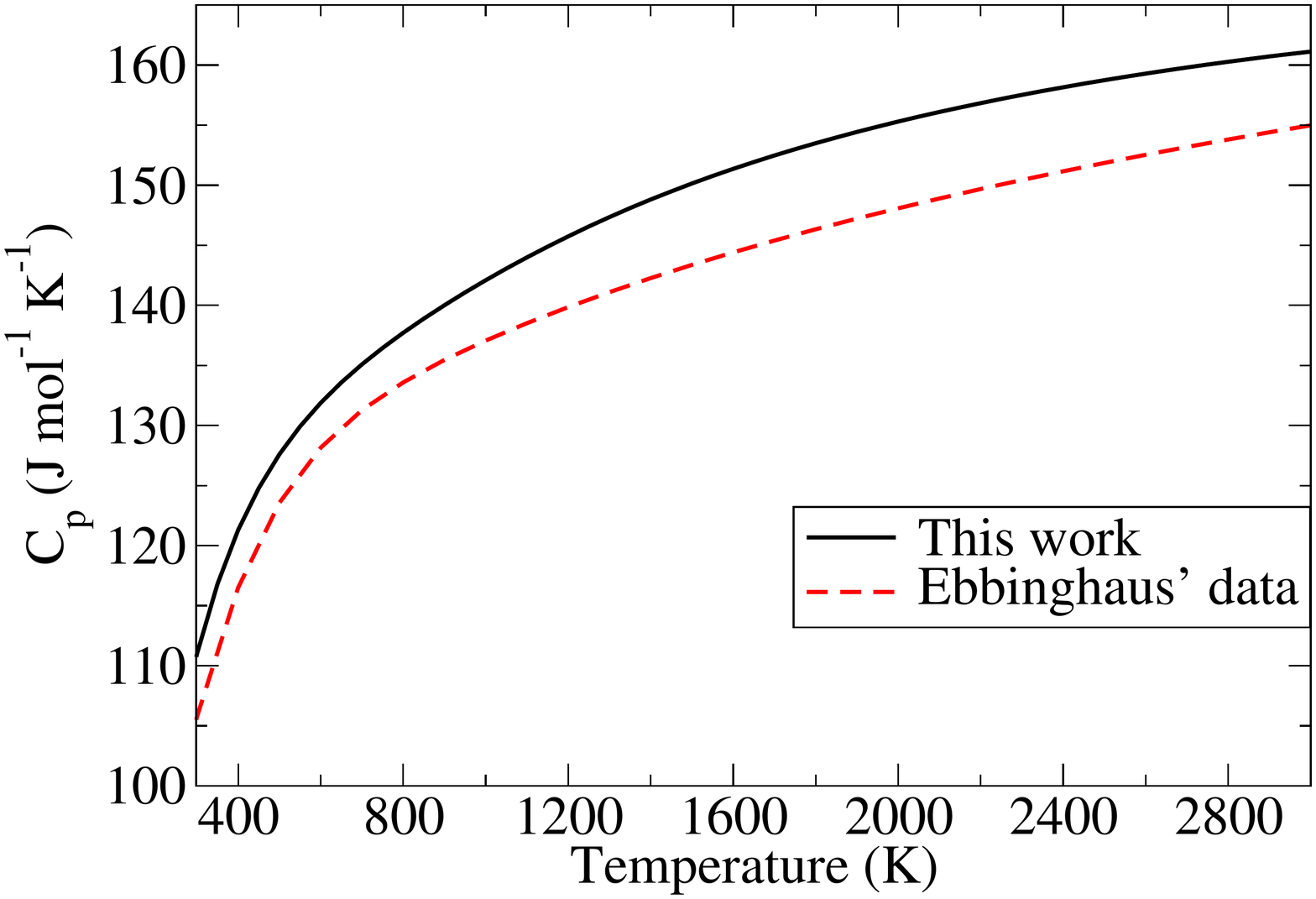}
\caption{Heat capacity of \ce{PuO2(OH)2(g)}. Ebbinghaus' data refers to Ref. \cite{actinide-Ebbinghaus-TR-2002}.}
\label{Fig:molPuH2O4}
\end{figure}

\clearpage
\bibliography{ms-PuOx.bib}